\documentclass[aps,prb,twocolumn,superscriptaddress,letterpaper]{revtex4}

\usepackage{etex} 
\usepackage{amsmath}
\usepackage{amssymb}
\usepackage{dsfont}
\usepackage[all,arc,knot,web]{xy}
\SelectTips{lu}{12}
\usepackage{graphicx}
\usepackage{hyperref}
\usepackage{color}
\usepackage{xcolor}
\definecolor{Blue}{rgb}{0.3,0.3,1}
\usepackage{pgffor}
\usepackage{verbatim}
\usepackage{tikz} 
\usepackage{wrapfig}

\usetikzlibrary{arrows,matrix,decorations.markings}
\usetikzlibrary{positioning}
\usetikzlibrary{decorations.markings}
\usetikzlibrary{decorations.pathmorphing}
\tikzset{->-/.style={decoration={markings,mark=at position #1 with {\arrow{>}}},postaction={decorate}}}
\tikzset{-<-/.style={decoration={markings,mark=at position #1 with {\arrow{<}}},postaction={decorate}}}
\newcommand{\unit}{\mathbf{1}}

\newcommand{\ket}[1]{\left|{#1}\right\rangle}
\newcommand{\bra}[1]{\left\langle{#1}\right|}
\newcommand{\bmm}{\begin{matrix}}
	\newcommand{\emm}{\end{matrix}}
\newcommand{\upperRomannumeral}[1]{\uppercase\expandafter{\romannumeral#1}}

\newcommand{\tTr}{\text{tTr}}

\definecolor{amethyst}{rgb}{0.6, 0.4, 0.8}

\newcommand{\TriangleYpart}[6]{
	\def\jjA{#1}
	\def\jjB{#2}
	\def\jjC{#3}
	\def\jjD{#4}
	\def\jjE{#5}
	\def\jjF{#6}
	\TriangleYpartExtended
}
\newcommand{\TriangleYpartExtended}[6]{
	\bmm\xy
	0;/r0.12pc/:;
	(-10,10)*{}="p1";
	(10,10)*{}="p2";
	(0,-10)*{}="p3";
	(-4,5)*{}="p4";
	(4,5)*{}="p5";
	(0,-3)*{}="p6";
	"p4";"p1" **\dir{-} ?(0.6)*\dir{#5}+(-3,-1) *{\scriptstyle \jjE};
	"p5";"p2" **\dir{-} ?(0.6)*\dir{#6}+(3.5,-1) *{\scriptstyle \jjF};
	"p3";"p6" **\dir{-} ?(0.6)*\dir{#1}+(3,-1) *{\scriptstyle \jjA};
	"p4";"p5" **\dir{-} ?(0.5)*\dir{#4}+(0,3.5) *{\scriptstyle \jjD};
	"p6";"p5" **\dir{-} ?(0.5)*\dir{#3}+(3.5,0) *{\scriptstyle \jjC};
	"p6";"p4" **\dir{-} ?(0.5)*\dir{#2}+(-3,0) *{\scriptstyle \jjB};
	\endxy\emm
}

\makeindex

\begin{document}
	
\title{The structure of fixed-point tensor network states characterizes patterns of long-range entanglement}
\author{Zhu-Xi Luo}
\author{Ethan Lake}
\affiliation{Department of Physics and Astronomy, University of Utah, Salt Lake City, Utah, 84112, U.S.A.}
\author{Yong-Shi Wu}
\affiliation{Key State Laboratory of Surface Physics, Department of Physics and Center for Field Theory and Particle Physics, Fudan University, Shanghai 200433, China}
\affiliation{Collaborative Innovation Center of Advanced Microstructures, Nanjing 210093, China} 
\affiliation{Department of Physics and Astronomy, University of Utah, Salt Lake City, Utah, 84112, U.S.A.}
\date{\today}

\begin{abstract}
The algebraic structure of representation theory naturally arises from 2D fixed-point tensor network states, which conceptually formulates the pattern of long-range entanglement realized in such states. In 3D, the same underlying structure is also shared by Turaev-Viro state-sum topological quantum field theory (TQFT). We show that a 2D fixed-point tensor network state arises naturally on the boundary of the 3D manifold on which the TQFT is defined, and the fact that exactly the same information is needed to construct either the tensor network or the TQFT is made explicit in a form of holography. Furthermore, the entanglement of the fixed-point states leads to an emergence of pre-geometry in the 3D TQFT bulk. We further extend these ideas to the case where an additional global onsite unitary symmetry is imposed on the tensor network states.
\end{abstract}


\maketitle

\section{Introduction}\label{Intro}
By now it is widely accepted that topological phases originate from the long-range entanglement existing in the condense matter system[\onlinecite{Chen2010}]. Tensor networks[\onlinecite{OrdusReview,Meets}], which focus on the wave functions of the system instead of the Hamiltonian, are generally considered as a natural tool to capture the behavior of long-range properties in a local way. The most successful examples of tensor network states include the Matrix Product States (MPS)[\onlinecite{MPS1991,MPS1992,MPS1993}] in 1D and the related Projected Entangled Pair States[\onlinecite{PEPS}] in 2D, both of which serve as an efficient ansatz for ground states of topological phases in their respective dimensions.

Besides its popularity in studying strongly-correlated systems, entanglement and tensor networks have also attracted increasing attention from the high energy theory community, in various attempts of realizing[\onlinecite{Swingle,Evenbly2011,QiEHM,HaPPY,YZ2015,YZ2016}] the holographic[\onlinecite{Hooft, Susskind}] AdS/CFT correspondence[\onlinecite{Maldacena}], and serving as a framework for Loop Quantum Gravity[\onlinecite{Rovelli1995,Rovelli1998,Barrett1998, Han2016}], both under the spirit of ``geometry from entanglement''. It is thus of theoretical interest to better understand the structure underlying tensor networks of topological phases, so as to formulate a more definitive theoretical framework for describing quantum entanglement. This is the subject we are concerned about in this paper, exemplified with a description of entanglement patterns in 2D topological phases. 

A tensor network is built from graphs consisting of interconnected tensors, imitating the structure of discrete lattices. The geometry of the network is generated by the pattern of interactions, namely, two sites in the network are close to each other if and only if they are entangled. Every tensor living on the sites of the network can be understood as a building block of entanglement. 

\begin{figure}[htbp]
	\centering
	\includegraphics[scale=1]{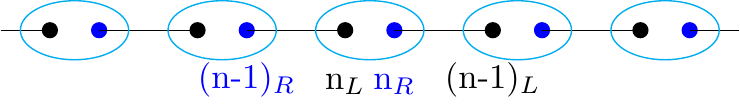}
	\caption{A parton construction of the AKLT state (see text for details).}
	\label{fig:AKLT}
\end{figure}

To illustrate, consider for example 
the celebrated AKLT states[\onlinecite{AKLT}] in a spin-$1$ chain. Such a state can be obtained from a parton construction.
As in Fig.\ref{fig:AKLT}, one regards every spin-1 degrees of freedom on site $n$ as a composite object consisting of two spin-1/2's at $n_L$ and $n_R$, and links each spin-1/2 spin on the site $n_L$($n_R$) to its nearest neighbor on $(n-1)_{R}$ ($(n+1)_L$) with a singlet bond. One then projects into the physical subspace with a spin-1 degree of freedom at each site. From the perspective of representation theory, the two spin-1/2's $n_L$ and $n_R$ can be combined as $\frac{1}{2}\otimes\frac{1}{2}=0\oplus 1$. The operator $P$ which projects into the physical subspace annihilates the first term on the right hand side and keeps only the spin-1 representation:
\begin{equation}
P:~\frac{1}{2}\otimes\frac{1}{2}\rightarrow 1.
\label{eq:Proj1}
\end{equation}

The tensor network representation of the AKLT state consists of tensors $T^i_{\mu\nu}$ at every site of the lattice, where the index $i\in\{0,\pm 1\}$ labels the physical spin-1 degrees of freedom on site $n$, while $\mu,\nu\in \{\uparrow,\downarrow\}$ label the auxiliary degrees of freedom associated with spin-1/2 partons at $n_L$ and $n_R$. The tensor $T$ can thus be understood as an adjoint map $T=P^\dagger$. 

The fusion algebra  $\frac{1}{2}\otimes\frac{1}{2}=0\oplus 1$ is a realization of entanglement, in the sense that after the fusion, the quantum states can no longer be factored as a product state of its local constituents (the spin-1/2 partons). This is manifested in the tensor network states in two ways: (i) when viewed as the adjoint of the projection $P$, the tensor $T^i_{\mu\nu}$ provides a way to encode the entanglement between $n_L$ and $n_R$; (ii) contraction of tensors on neighboring sites introduces maximal entanglement across the adjacent sites $n_L$ and $(n-1)_R$ (or $n_R$ and $(n+1)_L$).

In two dimensions, there exist intrinsic topological orders not protected by any symmetry. Following the discussion above, one would expect that in order to encode the entanglement of the topological phase in a tensor network, the structure of the tensors should be similar to that of a representation theory. Namely, it should incorporate the fusion the fusion algebra. For an intrinsic topological phase we further require the entanglement to be long-range. The pattern of long-range entanglement is captured by the fixed-point tensor network states that are invariant under renormalization group (RG) transformations of the tensor network. 

One natural question thus arises: for a general tensor network state to capture the long-range physics and to be a RG fixed-point state, what are the constraints that need to be satisfied? It turns out that in 2D, the input data for the tensor network indeed form a representation theory, in the form of a \textit{unitary fusion category}. In Sec.\ref{Sec:Cat} we briefly review the algebraic definition of a unitary fusion category (UFC). In Sec.\ref{Sec:Algebraic}, we discuss how the UFC structure arises from the fixed-point properties of general triple-line tensor network states, and conversely, how to generate fixed-point tensor network states from a UFC.

In Sec.\ref{Sec:Geo} we provide a geometrical point-of-view on the structure of fixed-point tensor network states by appealing to 3D state-sum topological quantum field theory (TQFT). The long-range physics of a phase with topological order is described by a TQFT[\onlinecite{Atiyah}]. The state-sum construction[\onlinecite{TV,Barrett1996}] of TQFTs discretizes the underlying manifold into a ``lattice'' or ``graph'', making explicit the locality of the theory. 

Recently the correspondence between 1D fixed-point tenor network states and 2D state-sum TQFT has been formulated rigorously[\onlinecite{Kapustin1607, Ryu1607}]. 
In one dimension higher, the state-sum construction of 3D TQFT was proposed by Turaev and Viro[\onlinecite{TV}] and later generalized by Barrett and Westbury[\onlinecite{Barrett1996}], which requires a UFC $\mathcal{C}$ as the input data. Given the UFC input data and a triangulation of the 3D manifold $\Sigma$, one can combinatorially define a topological invariant $\tau_{\mathcal{C}}(\Sigma)$ that is independent of the specific triangulation. 

When the underlying manifold $\Sigma$ contains boundaries, the TQFT can be viewed as a holographic map from its 3D bulk to the 2D boundary. Upon taking the Poincar\'e duality, this map produces the desired fixed-point tensor network state. Conversely, starting from a 2D fixed-point tensor network state, we show the generation of pre-geometries for the 3D bulk. It is a pre-geometry in the graph-theoretical sense: it contains vertices that correspond to points in the spacetime and oriented edges connecting them, i.e. a specific triangulation. The stronger concept of emergent bulk \textit{geometry} would further require the definition of a metric from entanglement measures in the tensor network. 

In Section \ref{Sec:Symmetry} we extend the framework to symmetric fixed-point tensor network states[\onlinecite{Singh2010, Ran2015}], which possesses a global onsite, finite, and unitary symmetry $\mathcal{G}$. The algebraic structure of these theories is given by $\mathcal{G}$-extension of the UFC $\mathcal{C}$, while the pre-geometric structure is closely related to 3D Homotopy Quantum Field Theory[\onlinecite{Homotopy2D, Homotopy3D, TuraevBook}].  
The construction is parallel to that of symmetry-enriched string-net models[\onlinecite{Levin1606, Cheng1606}].

\section{Review of unitary fusion categories}\label{Sec:Cat}

To prepare for the discussion of the algebraic structure of fixed-point tensor network states, in this section we briefly review the concept of a unitary fusion category (UFC).

A UFC $\mathcal{C}$ is a set of data $\{I,d,N,G\}$ subject to some consistency conditions. $I$ is the set of (isomorphism classes of) simple objects in $\mathcal{C}$. We require a trivial object $0\in I$. For every $j\in I$ there is a number $d_j\in\mathbb{R}$ called the quantum dimension of $j$, with $d_0=1$. The rank-3 tensor $N_{ijk}$ is a non-negative integer and describes the fusion rules between the objects $i,j$ and $k$. More specifically, the direct sum decomposition of the tensor product $i\otimes j\otimes k$ will include $N_{ijk}$ times the trivial object $0$. It is this feature of tensor product decomposition that gives UFC the interpretation of a representation theory. We assume multiplicity-free fusion rules throughout the paper, which means that we restrict to the case of $N_{ijk}\in\{0,1\}$ for $\forall i,j,k\in I$. We are also led to define the dual object $j^*$ as the only object that realizes $N_{0jj^*}=N_{jj^*0}=N_{j^*0j}=1$. It satisfies $j^{**}=j$ and $d_j=d_{j^*}$. Finally, to every six objects $i,j,k,l,m,n\in I$ we assign a quantum $6j$-symbol, which is a rank-6 tensor $G^{ijm}_{kln}\in \mathbb{C}$ (Relaxation of the multiplicity-free assumption would lead to four additional indices for the $G$-symbols). We will assume full tetrahedral symmetry of the $G$-tensors:
\begin{equation}\label{eq:UFC-Tetrahedral}
G^{ijm}_{kln}=G^{mij}_{nk^{*}l^{*}}=G^{klm^{*}}_{ijn^{*}}=\alpha_m\alpha_n\,\overline{G^{j^*i^*m^*}_{l^*k^*n}}.
\end{equation}
The number $\alpha_j$ is the Frobenius-Schur indicator:  $\alpha_j=sgn(d_j)$. The three equal signs correspond to the three generators of the $S_4$ symmetric (or tetrahedral) group, thus the name tetrahedral symmetry. Relaxing this condition leads to additional phase factors in the above equation that are the second or third roots of unity.  In the case where the input data are finite groups, the relaxation of tetrahedral symmetry can lead to the Dijkgraaf-Witten construction[\onlinecite{Dijkgraaf-Witten}] and the twisted quantum double model[\onlinecite{TQD}] based on three-cocycles, where time-reversal or(and) parity symmetry can generically be broken. Dropping the multiplicity-free condition and relaxing tetrahedral symmetry would complicate the problem, but we expect the main features of the correspondence to remain qualitatively the same.

For $\mathcal{C}$ to be a UFC, the above tensors need to satisfy certain consistency conditions, including:
\begin{itemize}
	\item[(UFC1)] Compatibility of $d_j$ and $N_{ijk}$:\begin{equation}
	d_i d_j = \sum_k N_{ijk^*} d_k.
	\end{equation}
	\item[(UFC2)] Pentagon equation:
	\begin{equation}\label{eq:UFC-Pentagon}
	\sum_{n}{d_{n}}G^{mlq}_{kp^{*}n}G^{jip}_{mns^{*}}G^{js^{*}n}_{lkr^{*}}=G^{jip}_{q^{*}kr^{*}}G^{riq^{*}}_{mls^{*}}.
	\end{equation}
	\item[(UFC3)] Orthogonality:
	\begin{equation}\label{eq:UFC-Ortho}
	\sum_{n}{d_{n}}G^{mlq}_{kp^{*}n}G^{l^{*}m^{*}i^{*}}_{pk^{*}n}=\frac{\delta_{iq}}{d_{i}}N_{mlq}N_{k^{*}ip}.
	\end{equation}
\end{itemize}
A useful identity that can be derived from above axioms is 
\begin{equation}
G^{ijk}_{0kj}v_jv_k=N_{ijk}.
\label{eq:FusionRuleIdentity}
\end{equation}

A UFC naturally arises from the representation theory of a finite group $\mathcal{G}$, i.e. $\mathcal{C}=Rep(\mathcal{G})$. The elements in the label set $I$ correspond to irreducible representations of $G$, and the $N_{ijk}$-tensor corresponds to the multiplicity of the representation $k^*$ in the direct sum decomposition of the tensor product $i\otimes j$. The $G$-tensors are simply the Racah $6j$ symbols of the group representation. 
More generally, a UFC is the representation category of a $C^*$-weak Hopf algebra.

\section{Algebraic Structure of Fixed-point Tensor Network States}\label{Sec:Algebraic}
In this section, we demonstrate the correspondence between fixed-point tensor network states and UFCs from an algebraic point of view. Part \ref{SubSec:CatLeft} derives the structure of the category from the fixed-point property of tensor network states, while part \ref{SubSec:CatRight} deals with the converse.

\subsection{Fixed-Point Tensor Network States Give Rise to a UFC}\label{SubSec:CatLeft}

\begin{figure}[htbp]
	\centering
	\includegraphics[]{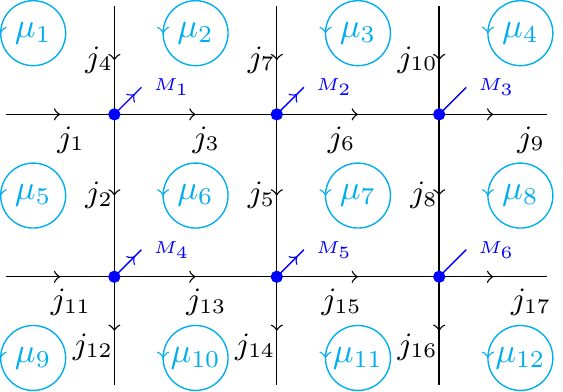}
	\caption{Example of a 2D tensor network (color online). The tensor living on the upper left vertex has components $(T^{M_1})^{j_1j_2j_3j_4}_{\mu_1\mu_5\mu_6\mu_2}$. The corresponding tensor network state of this graph is Eq.\eqref{eq:ExamplePEPS}.}
	\label{fig:ExamplePEPS}
\end{figure}

An example of a general 2D tensor network is displayed in Fig.\ref{fig:ExamplePEPS}. The tensors living on the vertices of the graph have one physical index $M$ that extends into the third dimension (out of the paper), as well as $4+4$  indices that correspond to internal degrees of freedom living on the links ($j$'s) and plaquettes ($\mu$'s) of the graph, which are auxiliary and are to be summed over. Generally the vertices in the network can be of valence $n$, with each tensor possessing $2n+1$ total indices. 

The tensor network state of Fig.\ref{fig:ExamplePEPS} is
\begin{equation}
\begin{split}
\ket{\Psi}= \sum\limits_{\{M_a\}} & \tTr\left[
(T^{M_1})^{j_1j_2j_3j_4}_{\mu_1\mu_5\mu_6\mu_2}\otimes
(T^{M_2})^{j_3j_5j_6j_7}_{\mu_2\mu_6\mu_7\mu_3}\right.\\
& \left.
\otimes (T^{M_3})^{j_6j_8j_9j_{10}}_{\mu_3\mu_7\mu_8\mu_4}\cdot\dots
\right]\ket{M_1,M_2,M_3,\cdots},\\
\end{split}
\label{eq:ExamplePEPS}
\end{equation}
where the tensor trace $\tTr$ indicates that all the internal indices $\{j_i\}$ and $\{\mu_i\}$ are contracted. Note that the tensor network commonly used is the special case where all auxiliary degrees of freedom $\mu$'s that live on the plaquettes are taken to be trivial.

To discuss the properties of fixed-point tensor network states, we work on a trivalent graph, or more specifically a honeycomb lattice which is bipartite and has $A,B$ sublattices. More general graphs can be easily obtained from trivalent graphs. 

\begin{figure}[htbp]
	\centering
	\includegraphics[]{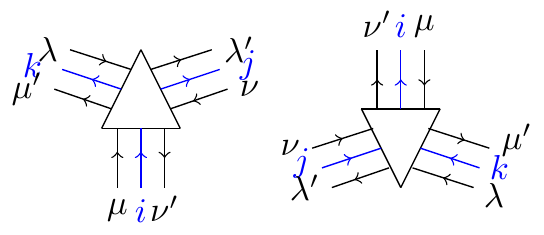}
	\caption{(Color online.) Triple-line structure of vertices that belong to $A$(left) and $B$(right) sublattices. For each of the three links that originally connected to some vertex, we sandwich it by two additional links. Upon projecting to the configurations that satisfy $\mu=\mu'=\cdots$, $\nu=\nu'=\cdots$, etc., one can see the plaquette degrees of freedom are restored. For simplicity, the physical indices $\{M_i\}$ are compressed in the figure.}
	\label{fig:AB-Vidal}
\end{figure}

Assign labels $i,j,k,\cdots\in I$ to every oriented link of the tensor network graph. For every $j\in I$ labeling some link, reversing the orientation of the link replaces $j$ with a dual label $j^* \in I$. We require the existence of an identity label $0 = 0^*$ in $I$. Associate labels $\mu,\nu,\lambda,\cdots\in I$ to each plaquette of the graph. 
These degrees of freedom are ``nonlocal'', in the sense that they can only be seen when looking at entire plaquettes. To encapsulate them in a strictly local way, we expand the above construction into a triple-line structure, following a procedure similar to that in Refs.[\onlinecite{Gu2009,Buerschaper2009}]. As shown in Fig.\ref{fig:AB-Vidal}, for each of the three links that originally connected to some vertex, we sandwich it between two additional links (the physical indices $\{M_i\}$ are suppressed for simplicity). Upon projecting to the configurations that satisfy $\mu=\mu'=\cdots$, $\nu=\nu'=\cdots$, etc., one can see the plaquette degrees of freedom are restored.

To construct a tensor network state, we assign a physical index $M=(i,j,k)$ to each vertex. The tensor on the corresponding vertex reads $(T^M)^{ijk}_{\mu\mu'\nu\nu'\lambda\lambda'}$. Here the superscripts $i,j,k$ are labels of the links joining at the specific vertex, while subscripts $\mu,\nu,\lambda$ are labels of the plaquette degrees of freedom adjacent to the vertex. (We stick to the rotationally invariant tensor network, where the permutations of the subscripts in $T_{jjj}$ and $T_{0jj^*}$ don't introduce extra phases. This property is related to the Tetrahedral symmetry[\onlinecite{Lan2014}] of the $6j$-symbols in the corresponding UFC.)

In 1D, RG flow corresponds to performing scale transformations by combining two or more adjacent tensors into one composite tensor. In 2D, RG transformations for tensor networks have been worked out in Refs.[\onlinecite{TERG,Evenbly2015,Evenbly2016}] in an approximate way. Exact invariance of tensor network states under RG flow in 2D can be regarded as an invariance of the tensor network state under 2D dual Pachner moves (see Fig.\ref{fig:2DDualPachner} below). These moves are discrete versions of diffeomorphisms of the underlying manifold. In Ref.[\onlinecite{Chen2010}], the authors discussed similar properties of fixed-point wave functions where the degrees of freedom live on the links of a network. The general situation where plaquette degrees of freedom are taken into consideration follows in a parallel way.

\begin{figure}[htbp]
	\centering
	\includegraphics[]{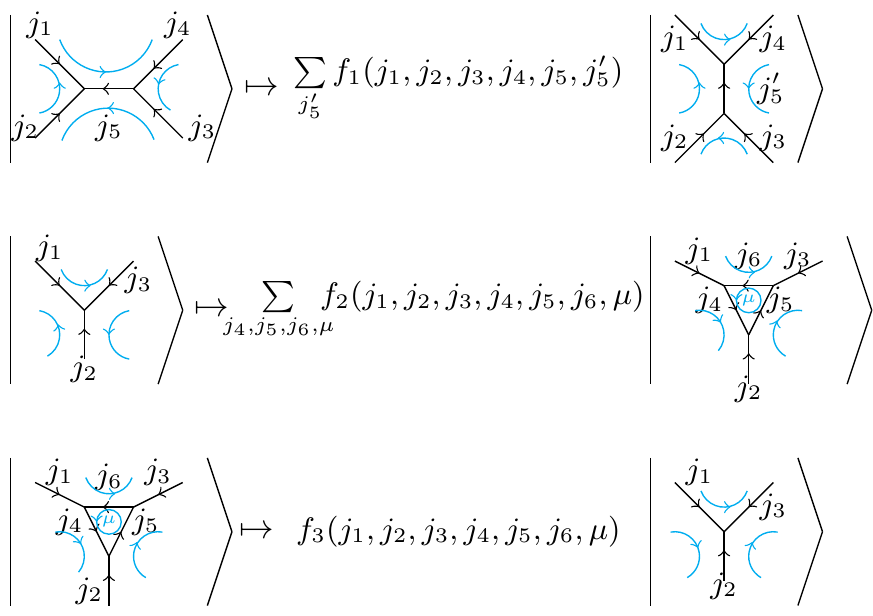}
	\caption{(Color online.) Dual Pachner moves in 2D. The top one is the $2-2$ recoupling move, while the second and third ones are the $1-3$ and $3-1$ moves. Since the dimension of Hilbert spaces is generically changed during the moves, we do not combine the latter two as is usually done in mathematical literature. The physical indices $\{M_a\}$ are understood to be attached to the vertices and thus suppressed in the figure.}
	\label{fig:2DDualPachner}
\end{figure}
Each Pachner move induces a linear transformation between the Hilbert spaces of different graphs, characterized by the coefficients $f_1,f_2,f_3$. The top move, denoted as $O_1$, is the 2-2 recoupling move, while the second and third ones $O_2, O_3$ are the 1-3 and 3-1 moves. Since the size of Hilbert spaces are changed during the moves, we do not combine the latter two as is usually done in mathematical literature.

The physical motivation for considering these moves comes from the fact that we are interested only in long-range physics. The two diagrams involved in the 2-2 recoupling move, when viewed from far away, both appear as a single four-valent vertex. If our tensor-network state is a fixed-point one, the two ways of decomposing this four-valent vertex into two three-valent vertices (by singular value decomposition) should be essentially the same, differing from one another only by a unitary transformation. The latter two 3$\leftrightarrow$1 moves correspond to usual local scale transformations of the graph, which allows us to take a zoomed-out view of the tensor network. 

Note that in the $2-2$ move, the plaquettes degree of freedom (colored cyan) are not changed. However, in the 1$\rightarrow$3 (3$\rightarrow $1) move, an additional closed string $\mu$ is added (removed) from the configuration. Consequently, while $f_1$ have no dependence on the plaquette strings, $f_2$ and $f_3$ do include $\mu$ as a nontrivial parameter.

In order for a tensor network state to be invariant under the Pachner moves, we require the following two necessary conditions:
\begin{itemize}
	\item[(C1)] The moves should be norm-preserving in the ground-state subspace. If $\ket{\Psi'}=O_i\ket{\Psi}$, then $\bra{\Psi'}\Psi'\rangle=\bra{\Psi}O_i^\dagger O_i\ket{\Psi}=\bra{\Psi}\Psi \rangle$. We emphasize that $\Psi'$ and $\Psi$ are not in the same Hilbert space, and that $O_i$ may not square matrices, i.e., the inverse matrices are not defined. 
	If one rotates the graph by $90$ degrees, then $O_1^\dagger$ can again be viewed as a $O_1$. The norm-preserving constraint then reads
	\begin{equation}\label{eq:P1}
	\left(O{O}_1^\dagger O{O}_1\right)=\unit,\ \left(O{O}_2O{O}_3\right)=\unit,\ \left(O{O}_3O{O}_2\right)=\unit,
	\end{equation}
	where the $\unit$ are identity matrices (of different dimensions).
	\item[(C2)] Two sequences of moves that result in the final tensor network configuration should be equivalent. If a final graph labeling $\{j_1',j_2',\cdots\}$ is obtained from some initial labeling  $\{j_1,j_2,\cdots\}$ through two (or more different sequences of Pachner moves, then we require the set of tensors $T_1(j_1',j_2',\cdots)$ and $T_2(j_1',j_2',\cdots)$ on each final graph configuration to be the same.
	\begin{equation}\label{eq:P2}
	O{O}_{\alpha_1}O{O}_{\beta_1}O{O}_{\gamma_1}\cdots=O{O}_{\alpha_2}O{O}_{\beta_2}\cdots.
	\end{equation}
\end{itemize}

These two conditions constrain the form of the functions $f_1,f_2,f_3$ in above Fig.\ref{fig:2DDualPachner}. From the first equation in \eqref{eq:P1}, one can derive, in terms of components, 
\begin{equation}
\delta_{j_5j_5''^*}=\sum_{j_5j_5''} f_1(j_4,j_1,j_2,j_3,j_5',j_5'')f_1(j_1,j_2,j_3,j_4,j_5,j_5').
\end{equation}
Similar formulas can be obtained for the other two equations in \eqref{eq:P1}.

Now we turn to condition $(P2)$, and construct commutative diagrams  from sequences of $O{O}$ operators that result in identical tensor network configurations. Requiring these diagrams to commute will allow us to place various consistency conditions on the $f_i$ matrix elements. For tensor network configurations with two and three uncontracted legs, there are no nontrivial commutative diagrams. For tensor network configurations with four uncontracted legs, the only operations we are allowed to do are already fully captured by $O O_2$ and $O O_3$. But constraints do arise for commutative diagrams involving tensor networks with five uncontracted legs. Indeed, choose the two sequences below in Fig.\ref{fig:Pentagon}.

\begin{figure}[htbp]
	\centering
	\includegraphics[]{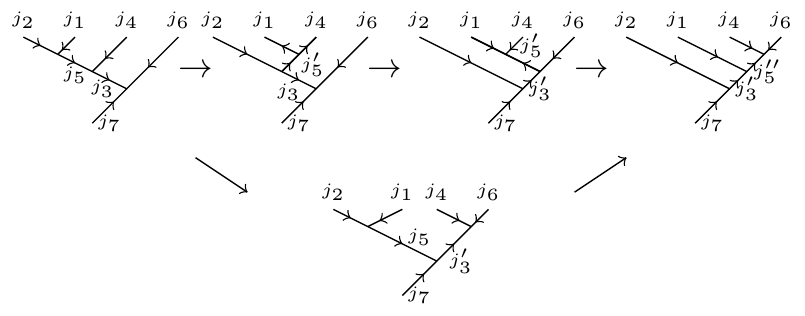}
	\caption{The two different sequences of Pachner moves that share the same initial and final configurations.}
	\label{fig:Pentagon}
\end{figure}
The constraint that the above diagram must commute leads to

{\small 
\begin{equation}
\begin{split}
	& \sum_{j_5'}f_1(j_1,j_2,j_3^*,j_4,j_5^*,j_5')f_1(j_5'^*,j_2,j_7,j_6,j_3^*,j_3')\\
	& ~~~~~ \times f_1(j_4,j_1,j_3',j_6,j_5'^*,j_5'')\\ 
	=~ & f_1(j_4,j_5,j_7,j_6,j_3,j_3')f_1(j_1,j_2,j_7,j_5'',j_5^*,j_3').\\
\end{split}
\end{equation}
}

Below we show that the functions $f_1, f_2, f_3$ are closely related to the $6j$-symbols ($G$-tensors) introduced in the previous section \ref{Sec:Cat}, from which we can reconstruct the fusion rules ($N$-tensors) and quantum dimensions ($d$'s), thereby arriving at a UFC.

We introduce a new set of symbols with six parameters by
\begin{equation}
\label{eq:ChangeVariables}
\frac{G^{ijm}_{kln}}{G^{m^*m0}_{00m}G^{n^*n0}_{00n}}=f_1(i,j,k,l,m,n),
\end{equation}
the above conditions (C1),(C2) reduce to
\begin{equation}\label{eq:P12Standard}
\begin{split}
\text{(P1):}\ &  \delta_{j_5j_5''^*}=\sum_{j_5j_5''}d_{j_5'}v_{j_5}v_{j_5''}G^{j_1j_2j_5}_{j_3j_4j_5'}G^{j_4j_1j_5'}_{j_2j_3j_5''}.\\
\text{(P2):}\ & \sum_{j_5'}d_{j_5'}G^{j_1j_2j_5^*}_{j^*_3j_4j_5'}G^{j_5'^*j_2j_3^*}_{j_7j_6j_3'}G^{j_4j_1j_5'^*}_{j_3'j_6j_5''}=G^{j_4j_5j_3}_{j_7j_6j_3'}G^{j_1j_2j_5^*}_{j_7j_5''j_3'}.
\end{split}
\end{equation}

Here we have defined 
\begin{equation}
v_j:=\frac{1}{G^{j^*j0}_{00j}},~~~d_j:=v_j^2.
\label{eq:QuantumDimension}
\end{equation}


Comparing the above equations with \eqref{eq:UFC-Pentagon} and \eqref{eq:UFC-Ortho}, we recognize that the norm-preservation condition on Pachner transformations recovers the orthogonality condition, while the path-independence of Pachner transformations recovers the pentagon condition in the definition of a UFC.
By appealing to the coherence theorem[\onlinecite{MacLane}], any commutative diagram involving two ways of relating two tensor network configurations with $n > 5$ uncontracted legs to one another will commute as long as the pentagon identity holds, and so the above conditions exhaust the constraints we can put on the $G$ tensors.

The two sequences in Fig.\ref{fig:Pentagon} involve only the $O{O}_1$ move. One can choose other sequences involving $1\leftrightarrow 3$ moves and derive the relationship between $f_1$, $f_2$ and $f_3$. They differ in prefactors by the product of powers of $d$'s and $D=\sum_j d_j^2$. We rewrite the equation for these Pachner moves in Eq.\eqref{eq:NewPachner}, in which the plaquette labels that do not change during the moves are suppressed.

\vspace*{0.4cm}

\begin{widetext}
	\begin{equation}
	\label{eq:NewPachner}
	\begin{split}
	\text{(P1)}:\ & (T^{M_1})^{j_1j_2j_5} (T^{M_2})^{j_5^*j_3j_4}= \sum_{j_5'} v_{j_5}v_{j_5'}G^{j_1j_2j_5}_{j_3j_4j_5'} (T^{M_3})^{j_1j_5'j_4} (T^{M_4})^{j_5'^*j_2j_3},\\
	\text{(P2)}:\ & 
	(T^{M})^{j_1j_2j_3}=\sum_{j_4,j_5,j_6,\mu} v_{j_4}v_{j_5}v_{j_6} G^{j_2j_3j_1}_{j_6^*j_4j_5^*}(T^{M_1})^{j_1j_4^*j_6}_\mu (T^{M_2})^{j_2j_5^*j_4}_\mu (T^{M_3})^{j_3j_6^*j_5}_\mu, \\
	\text{(P3)}:\ & \sum_{\mu} (T^{M_1})^{j_1j_4^*j_6}_\mu (T^{M_2})^{j_2j_5^*j_4}_\mu (T^{M_3})^{j_3j_6^*j_5}_\mu=\frac{v_{j_4}v_{j_5}v_{j_6}}{D} G^{j_3^*j_2^*j_1^*}_{j_4^*j_6j_5^*} (T^{M})^{j_1j_2j_3}. \\
	\end{split}
	\end{equation}
\end{widetext}

Tetrahedral symmetry \eqref{eq:UFC-Tetrahedral} is guaranteed by the rotational invariance of the graph, or equivalently by the permutation symmetry of the tensors $T_{ijk}=T_{kij}=T_{jki}$.

\begin{figure}[htbp]
	\centering
	\includegraphics[]{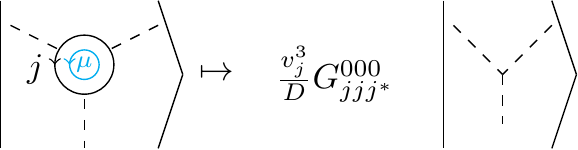}
	\caption{(Color online.) Taking $j_1=j_2=j_3=0$ for the 3-1 move in Fig.\ref{fig:2DDualPachner}, one recovers the quantum dimensions $d_j$.}
	\label{fig:QuantumDimension}
\end{figure}

To see the physical meaning of the definition in Eq.\eqref{eq:QuantumDimension}, we can take $j_1=j_2=j_3=0$ for the 3-1 move in Fig.\ref{fig:2DDualPachner}. The constraint $j_4=j_5=j_6$ must be satisfied, and so (suppressing the irrelevant plaquette degrees of freedom) the move simplifies as in Fig.\ref{fig:QuantumDimension}.

Using tetrahedral symmetry and Eq.\eqref{eq:QuantumDimension}, we see that  $v_j^3G^{000}_{jjj^*}=v_j^3G^{j^*j0}_{00j}=d_j$, consistent with the physical meaning of the quantum dimensions.

\begin{figure}[htbp]
	\centering
	\includegraphics[]{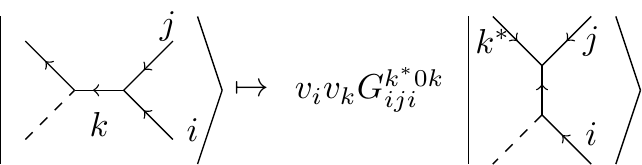}
	\caption{Taking $j_2=0$ in the 2-2 move in Fig\ref{fig:2DDualPachner}, one recovers the fusion rules of $N_{ijk^*}$. }
	\label{fig:FusionRule}
\end{figure}

The fusion rules are also encoded in the $G$-symbols, as can be observed already from Eq.\eqref{eq:FusionRuleIdentity}. When one takes $j_2=0$ in the 2-2 move, then $j_1=j_5^*$ and $j_3=j_5'$ must be satisfied. Rewriting $i=j_3, j=j_4$ and $k=j_5$, the move reduces to Fig.\ref{fig:FusionRule}.

Using Eq.\eqref{eq:FusionRuleIdentity}, we see that $v_{i}v_{k}G^{k^*0k}_{iji}=N_{ijk^*}$. Consequently, the tensor network configuration on the right is only allowed if $N_{ijk^*}$ is nonzero, i.e. if the branching rules are satisfied.

Combining the results above, we see that the fixed-point requirement of a tensor network state leads naturally to a set of data $\{I,d,N,G\}$ that satisfy the axioms of UFC $\mathcal{C}$.

\subsection{Construction of a Fixed-Point Tensor Network State from a UFC}\label{SubSec:CatRight}

Having shown how a fixed-point tensor network state contains the data of a UFC, we now show how one can begin with a UFC $\mathcal{C}$ and construct a fixed-point tensor network state. We will use a triple-line tensor network construction, and will color the triple-line structure by assigning the labels $i,j,k,\cdots\in I$ to the central (blue) links in Fig.\ref{fig:AB-Vidal} and the labels $\mu,\nu,\lambda,\cdots \in I$ to the adjacent black links as before. We organize the labellings in a way so that for any three central links $i,j,k$ that point to a common vertex, $N_{ijk}\neq 0$. 

\begin{figure}[htbp]
	\includegraphics[scale=.35]{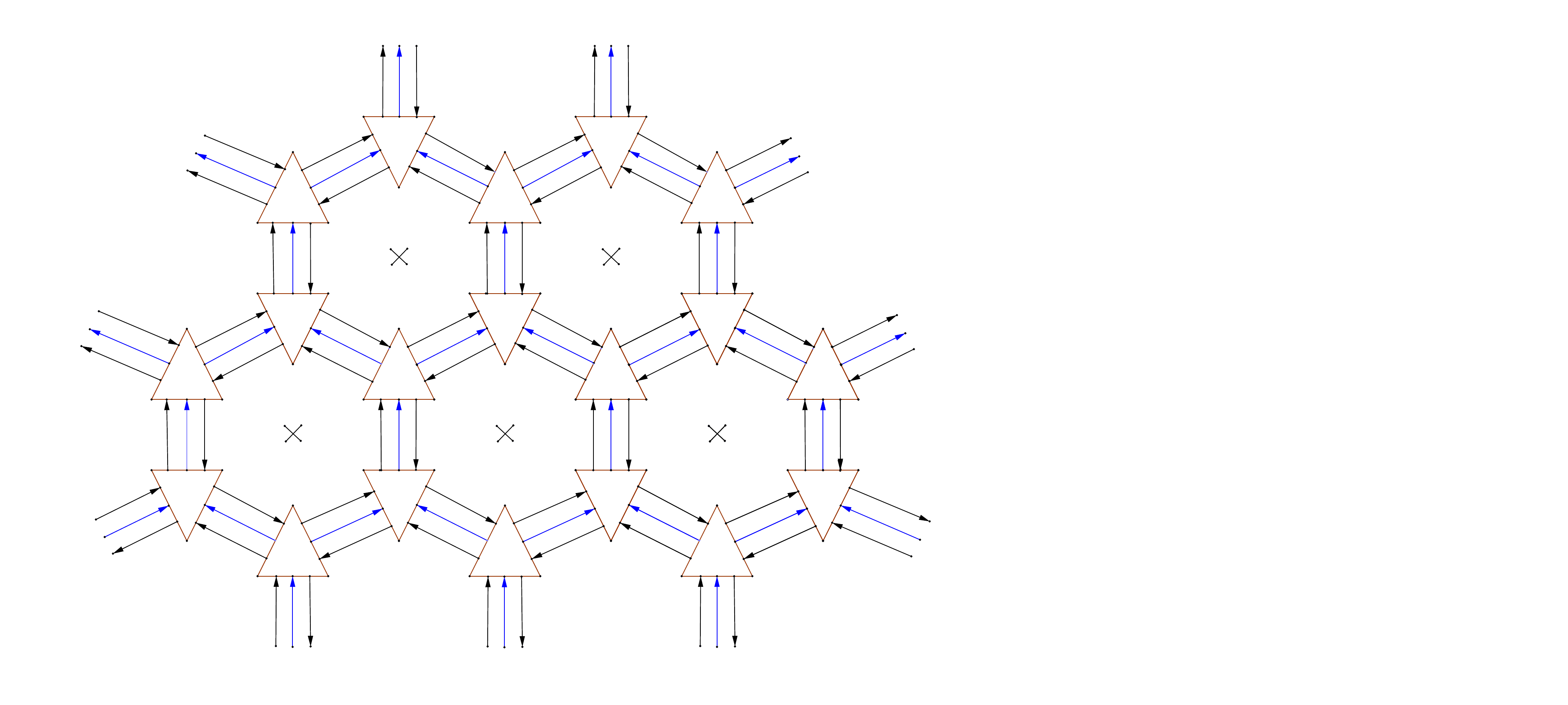}
	\caption{(Color online.) An illustration of the tensor network with tensors given by Eq. \eqref{eq:TN-Tensors}. The upward-pointing (downward-pointing) triangles are located on the A (B) sublattice.}
	\label{fig:Network-Vidal}
\end{figure}

The next step is to import the $\{G\}$-tensors from the UFC into the tensor network. For Fig.\ref{fig:AB-Vidal}, we associate to every vertex (small triangle) a tensor $T$ on the $A$ and $B$ sublattices in the following way (parallel to Refs.[\onlinecite{Gu2009,Buerschaper2009}]):
\begin{equation}\label{eq:TN-Tensors}
\begin{split}
& \text{A:}\ (T^{M})^{ijk}_{\mu\mu'\nu\nu'\lambda\lambda'}=\frac{\left(v_{\mu}v_{\nu}v_{\lambda}\right)^{1/3}}{\sqrt{D}}\sqrt{v_iv_jv_k}G^{ij^*k^*}_{\lambda\mu^*\nu}\delta_{\mu\mu'}\delta_{\nu\nu'}\delta_{\lambda\lambda'},\\
& \text{B:}\ (T^{M})^{ijk}_{\mu\mu'\nu\nu'\lambda\lambda'}=\frac{\left(v_{\mu}v_{\nu}v_{\lambda}\right)^{1/3}}{\sqrt{D}}\sqrt{v_iv_jv_k}G^{i^*jk}_{\lambda\mu^*\nu}\delta_{\mu\mu'}\delta_{\nu\nu'}\delta_{\lambda\lambda'},\\
\end{split}
\end{equation}
where we have denoted $v_j=\sqrt{d_j}$ for $j\in I$. The physical index $M$ is defined as the triple $(i,j,k)$. Upon contracting the internal indices as demonstrated in Fig.\ref{fig:Network-Vidal}, one arrives at a tensor network state.

It was proved in Ref.[\onlinecite{Gu2009}] that these states are fixed-point states under renormalization transformations of the tensor network. More precisely, they are invariant under the 2D dual Pachner moves of Fig. \ref{fig:2DDualPachner}. In the Hamiltonian language, these fixed-point states are the ground states of string-net models[\onlinecite{LW}]. 

Note that if the tensors appearing in Eq.\eqref{eq:TN-Tensors} are to be non-zero, we must have
\begin{equation}
\begin{split}
& \text{A:}~~N_{\mu^*i\nu}=N_{\nu\lambda^*j}=N_{\lambda\mu^*k}=N_{ij^*k^*}=1,\\
& \text{B:}~~N_{\mu^*i^*\nu}=N_{\nu\lambda^*j^*}=N_{\lambda\mu^*k^*}=N_{i^*jk}=1.
\end{split}
\end{equation}
Although the two sets $\mu,\nu,\lambda,\cdots$ and $i,j,k,\cdots$ both take values in the label set $I$ of the UFC $\mathcal{C}$, they are {\it not} on the same physical footing. The origin of the above form \eqref{eq:TN-Tensors} of tensors is the following. 

Denote $B_p^\mu$ for plaquette $p$ of the graph as the operator that adds a closed loop $\mu$ inside $p$. Further define 
\begin{equation}
B_p=\sum_\mu \frac{d_\mu}{D}B_p^\mu.
\label{eq:Bp}
\end{equation}
$B_p$ is the composition of the elementary $1-3$ and $3-1$ moves in \ref{fig:2DDualPachner}:
\begin{equation}
B_p=O_2\circ O_1.
\end{equation}
It changes the labellings of the links the surround the plaquette $p$ while keeping all the other labellings in the graph untouched. Since it is a composition of the elementary moves, it keeps the fixed-point tensor network state invariant. 

\begin{figure}[htbp]
	\centering
	\includegraphics[]{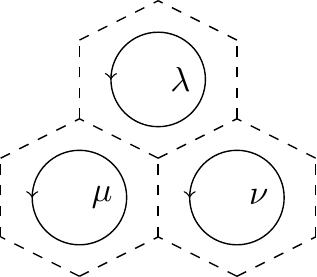}
	\caption{Construction of fixed-point tensor network state using loops.}
	\label{fig:LoopState}
\end{figure}

The tensor network state can actually be constructed from this operator as
\begin{equation}
\begin{split}
\ket{\Psi} & =\prod_p B_p \ket{0}\\
& = \sum_{\mu,\nu,\lambda,\dots}\frac{d_\mu}{D}\frac{d_\nu}{D}\frac{d_\lambda}{D}\cdots\ket{\mu,\nu,\lambda,\cdots}_{coh},\\
\end{split}
\label{eq:LoopState}
\end{equation}
where the state $\ket{0}$ means the graph is empty, i.e., we assign the vacuum string $0$ to every link to the graph, and $\ket{\mu,\nu,\lambda,\dots}_{coh}$ denotes the state with $\mu,\nu,\lambda,\dots$ as plaquette degrees of freedom and with all links carrying the label $0$. $\ket{\mu,\nu,\lambda,\cdots}_{coh}=B^\nu_{p_1}B^\mu_{p_2}B^\lambda_{p_3}\dots\ket{0}$ as demonstrated in the Fig.\ref{fig:LoopState}. The factor $d_\mu$ attributes to the fact that every closed string $\mu$ has an amplitude of $d_\mu$. Notice that the state $\ket{\mu,\nu,\lambda,\cdots}_{coh}$ are coherent states; they are not necessarily orthogonal. Furthermore, all the closed loops appearing in the state $\ket{\mu,\nu,\lambda,\cdots}_{coh}$ are independent of each other, i.e., mutually un-entangled.

Since they are fixed-point states, one can further translate the degrees of freedom from the loops to the links using dual 2D Pachner moves described in Fig.\ref{fig:2DDualPachner}. This gives
\begin{equation}\label{eq:TensorState}
\begin{split}
\ket{\Psi}=& \sum_{M_1,M_2,\cdots} \tTr\left[\otimes_v T^{M_v}\right] \ket{M_1,M_2,\cdots}.
\end{split}
\end{equation}
where the $\{M_i\}$ are physical indices.
The $T$ tensors appearing in the tensor trace take values exactly as in Eq.\eqref{eq:TN-Tensors}. 

The above procedure presents an analogy to the 1D AKLT example discussed in the introduction. The tensor network representation of the AKLT state encodes entanglement in two ways: (i) the spin-1/2 partons $\mu,\nu,\cdots$ on neighboring sites (e.g., $(n-1)_R$ and $n_L$) are entangled as singlets, and (ii) the partons on the same site (e.g., $n_L$ and $n_R$) as entangled as triplets. While the former is realized by the contraction of tensors, the latter entanglement is carried by every single tensor in the network.

Similarly in two dimensions, entanglement is created in several steps. (i)  In Eq.\eqref{eq:LoopState}, one first uses the $B_p$ operators to generate plaquette degrees of freedom $\mu,\nu,\lambda,\cdots$. In the language of triple-line structure, this corresponds to taking all the $i,j,k,\cdots=0$ in Fig.\ref{fig:AB-Vidal}, and contracting all the the partons $\mu=\mu'$ , $\nu=\nu'$ etc. The latter creates entanglement inside every plaquette. (ii) The next step is to project onto the physical degrees of freedom $i,j,k,\cdots$, which are defined on the links and correspond to the spin-1 degrees of freedom in the AKLT analogy. Entanglement is created when this projection takes place, i.e. when one uses dual Pachner moves to fuse the loops $\mu,\nu,\lambda,\cdots$ and rearrange the degrees of freedom from the plaquettes to the links. Mathematically, this is realized by the fusion $\delta$-tensors in the UFC. (iii) Finally, the $i,j,k,\cdots$ are contracted, resulting in the entanglement between different sites.

The pattern of entanglement in the second step manifests itself as $6j$ $G$-symbols in the coefficients generated by the Pachner moves, which become encoded in the tensors $T$ in Eq.\eqref{eq:TN-Tensors}. These local tensors record the history of the projection in step (ii) by representing the initial mutually {\it un-entangled} parton degrees of freedom in terms of the {\it entangled} physical degrees of freedom. If the label set of the UFC contains only one trivial object $I=\{0\}$ (and thus $D=1$), then the entanglement is short-range. Generally if one starts from a nontrivial UFC, the constructed fixed-point tensor networks state will be long-range entangled. We conclude that the local $T$-tensors are the building blocks of long-range entanglement in the corresponding topological phase.

\section{Geometrical Perspective of the Correspondence}\label{Sec:Geo}
In this section we discuss the geometric structure of fixed-point tensor network states and its relationship to a 3D Turaev-Viro state-sum TQFT. We briefly review a few basic TQFT facts. On a three dimensional manifold $\Sigma$, a full-extended unitary 3D TQFT is a symmetric monoidal functor[\onlinecite{Atiyah}] from the category of three-cobordisms to the category of vector spaces over $\mathbb{C}$:
\begin{equation}
\mathcal{F}:3Cob\rightarrow Vect_{\mathbb{C}}.
\end{equation}
Specifically, we assign a Hilbert space of states $\mathcal{H}$ to each spatial slice (2D manifold) of a three-cobordism. If the spatial slice contains a disjoint union of $n$ 2D manifolds, the corresponding Hilbert space splits through the tensor product as $\mathcal{H}^{\otimes n}$. A 3D TQFT associates to $\Sigma$ a linear map from $\mathcal{H}^{\otimes n_i}$ to $\mathcal{H}^{\otimes n_o}$, where $n_i$ is the number of disjoint parts of the incoming spatial slice, and $n_o$ the number for outgoing spatial slice.

The cylinder map is the identity $id: \mathcal{H}\rightarrow\mathcal{H}$. If the 3D manifold $\Sigma$ is closed, then the map is a partition function $Z(\Sigma):\mathbb{C}\rightarrow\mathbb{C}$. Other simple examples include the cap cobordism, where the map is $Tr: \mathcal{H}\rightarrow \mathbb{C}$; the cup cobordism $\eta:\mathbb{C}\rightarrow\mathcal{H}$; the product bordism (a pair of pants) $m: \mathcal{H}^{\otimes 2}\rightarrow \mathcal{H}$; and  the coproduct bordism (an inverted pair of pants) $\Delta: \mathcal{H}\rightarrow\mathcal{H}^{\otimes 2}$. 

\subsection{3D State Sum TQFT}\label{SubSec:GeoTQFT}

A state-sum construction of a TQFT is a discretization of the above formalism. The algebraic data needed to define a 3D state-sum TQFT form a UFC $\mathcal{C}$  in the following way. 

\begin{figure}[htbp]
\centering
\includegraphics[]{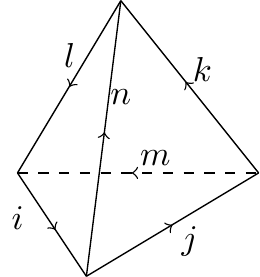}
\caption{Every tetrahedron is associated with a $G$-tensor. This specific configuration corresponds to $G^{ijm}_{kln}$.}
\label{fig:Tetrahedron6j}
\end{figure}

We start from a \textit{closed} three dimensional manifold $\Sigma$, and define on it a triangulation $\mathcal{T}(\Sigma)$. An oriented coloring of the triangulation refers to the assignment of a label $j\in I$ to every 1-simplex (edge) of the triangulation. Substituting $j$ by $j^*$ and reversing the arrow leaves the oriented coloring invariant. Then we associate[\onlinecite{TV,Barrett1996}] a tensor $G^{ijm}_{kln}$ to each tetrahedron with edges labeled by $\left\{i,j,m,k,l,n\right\}$, as indicated in Fig.\ref{fig:Tetrahedron6j}. The tetrahedral symmetry condition \eqref{eq:UFC-Tetrahedral} can be understood geometrically as the requirement that viewing the tetrahedron from four different directions give rise to the same tensor.

The weight of a specific coloring is a number defined as the product of all $G$ tensors for all tetrahedra in the triangulation and the product of all $d_j$ for all edges in the triangulation. The Turaev-Viro invariant for the manifold $\Sigma$ is then computed as the summation of these weights over all colorings of the triangulation $\mathcal{T}(\Sigma)$. Schematically, we have

\begin{equation}
\label{TV}
\tau_{\mathcal{C}}(\Sigma)=\sum_{labellings}\prod_{vertices}\frac{1}{D}\prod_{tetrahedron}G\prod_{edges}d,
\end{equation}
where the total quantum dimension $D=\sum\limits_{j\in I} d_j^2$.

\begin{figure}[htbp]
\centering
\includegraphics[]{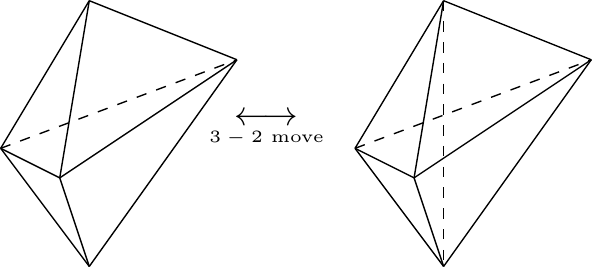}
\caption{3D Pachner 3-2 move in the triangulation picture.}
\label{fig:3DPachner3-2Original}
\end{figure}

\begin{figure}[htbp]
\centering
\includegraphics[]{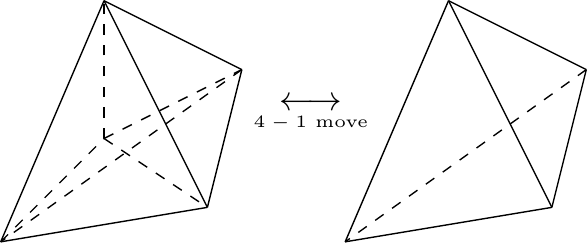}
\caption{3D Pachner 4-1 move in the triangulation picture.}
\label{fig:3DPachner4-1Original}
\end{figure}

Independence of the invariant $\tau_{\mathcal{C}}(\Sigma)$ with respect to triangulations of $\mathcal{T}(\Sigma)$ can be shown by following a standard procedure. Any two different triangulations in 3D can be related by a sequence of 3D Pachner moves[\onlinecite{Pachner}] depicted in Fig.\ref{fig:3DPachner3-2Original} and \ref{fig:3DPachner4-1Original}. Invariance of the state-sum under these moves corresponds exactly to the consistency condition (UFC2) and (UFC3) above, namely, the Pentagon equation and the Orthogonality condition. We demonstrate this correspondence in detail in Appendix \ref{App:Pachner}. Consequently, the input category $\mathcal{C}$ being a UFC automatically guarantees this topological invariance. 

\subsection{Manifolds With Boundary}\label{SubSec:GeoBdry}

The above discussion can be generalized to the case where $\Sigma$ has 2D boundaries $\partial \Sigma$[\onlinecite{Karowski1992}]. Following the notation of Ref.[\onlinecite{Kapustin1607}], we call the initial and final spatial slices of the cobordism as cut boundaries, and all others as brane boundaries. Cobordisms are composed along cut boundary, while boundary conditions need to be imposed on brane boundaries. 

\begin{figure}[htbp]
	\centering
    \includegraphics[]{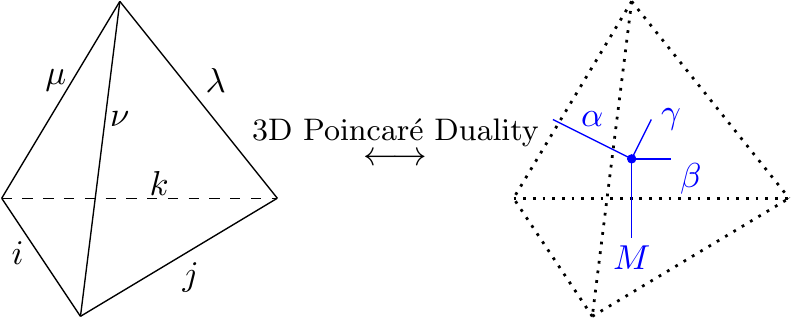}
	\caption{(Color online.) Poincar\'e Duality applied to a tetrahedron on the brane boundary of $\Sigma$.}
	\label{fig:3DRealPoincare}
\end{figure}

Consider the special case where $\partial \Sigma$ consists of one single component of both cut boundary and brane boundary. One tetrahedron $[i,j,k,\mu,\nu,\lambda]$ in the triangulation near the brane boundary is depicted in Fig.\ref{fig:3DRealPoincare}. The faces $(\mu,i,\nu)$, $(\nu,j,\lambda)$ and $(\lambda,k,\mu)$ lie on the brane boundary, while the face $(i,j,k)$ is in the bulk. There can be a large number of tetrahedra between the $(i,j,k)$ plane and the cut boundary(initial spatial slice), but one can use Pachner moves in Fig.\ref{fig:3DPachner3-2Original},\ref{fig:3DPachner4-1Original} to reduce the number of tetrahedra in the bulk and effectively arrive at a single ``layer'' of tetrahedra that looks like Fig.\ref{fig:3DRealPoincare}. In other words, without loss of generality, one can view $(i,j,k)$ as living on the cut boundary.

Applying Poincar\'e duality, we can associate 3-simplices (tetrahedra) with 0-simplices (vertices) of the dual graph, and 2-simplices (faces) with 1-simplices (edges) of the dual graph. In the tetrahedron $[i,j,k,\mu,\nu,\lambda]$, $\alpha$ is dual to the triangle bounded by the three links $(\mu,i,\nu)$, $\beta$ is dual to the triangle bounded by $(\nu,j,\lambda)$, $\gamma$ is dual to the triangle $(\lambda,k,\mu)$ and the index $M$ is the collection $(i,j,k)$.

The links $\alpha$ must match up with another link $\alpha'$ in the triangulation, which comes from the dual of another tetrahedron that shares the face $(\mu,i,\nu)$ with the above tetrahedron. A similar identification occurs for $\beta,\gamma$, etc.. Consequently, links carrying the indices $\alpha,\beta,\gamma,\dots$ form a 2D trivalent graph, with the extra links like $M$ dangling in the third dimension of this graph.

The graph generated by Poincar\'e duality in this way coincides exactly with the setup of a 2D tensor network. We see that the original edges $\mu,\nu,\lambda$ of the triangulation map to the plaquette degrees of freedom in the dual picture. This precisely gives rise to the triple-line structure depicted in Fig.\ref{fig:AB-Vidal}. 

The mapping from internal indices $\alpha,\beta,\gamma$ to the physical index $M$ can be interpreted as a boundary-to-bulk map in the TQFT context. The factors of $\delta_{\mu\mu'}\delta_{\nu\nu'}\delta_{\lambda\lambda'}$ in Eq.\eqref{eq:TN-Tensors} can now be understood as well: these are the constraints that ensure the plaquette degrees of freedom in the dual graph are associated to links in the original triangulation in a well-defined way. In other words, these constraints entangle the 2-simplices in the same tetrahedron.

Pachner moves in the original triangulation picture Fig.\ref{fig:3DPachner3-2Original} and \ref{fig:3DPachner4-1Original} map to the dual Pachner moves of Fig.\ref{fig:2DDualPachner}. This is related to the fact we mentioned above: both moves correspond algebraically to the Pentagon and Orthogonality axioms of 3D state-sum TQFTs.

\begin{figure}[htbp]
\centering
\includegraphics[scale=.55]{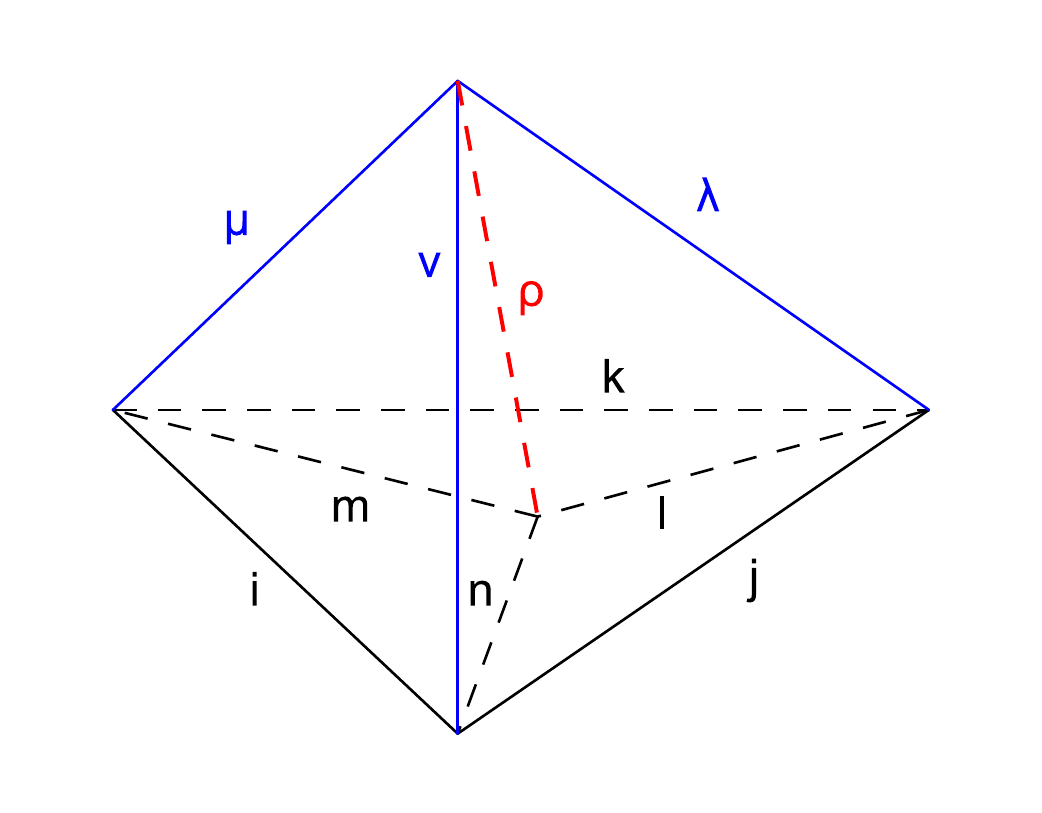}
\caption{(Color online.) Gluing of three tetrahedra.}
\label{fig:3Tetrahedra}
\end{figure}

\begin{figure}[htbp]
\centering
\includegraphics[scale=.6]{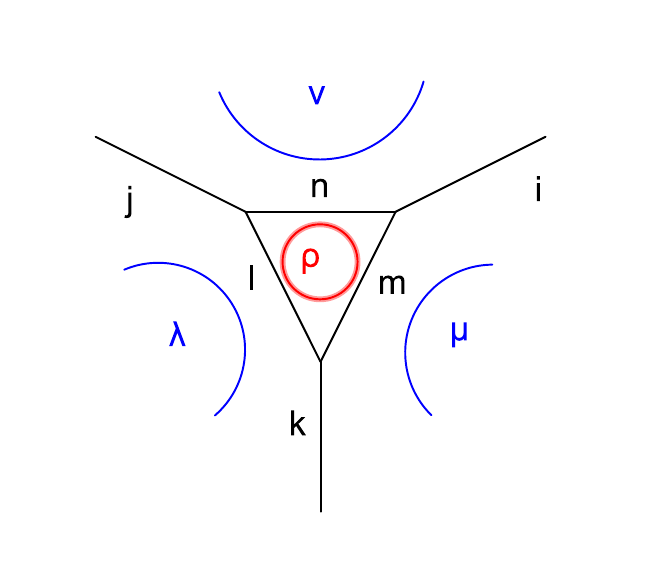}
\caption{(Color online.) Gluing of three tetrahedra corresponds to constructing a triangular plaquette.}
\label{fig:3Tetrahedra-dual}
\end{figure}

Consider the situation with three tetrahedra are glued together as in Fig.\ref{fig:3Tetrahedra}. In the tensor network picture (dual to the triangulation picture), this corresponds to the triple-line structure near a triangular plaquette (Fig.\ref{fig:3Tetrahedra-dual}).

\begin{figure}[htbp]
	\centering
	\includegraphics[scale=.5]{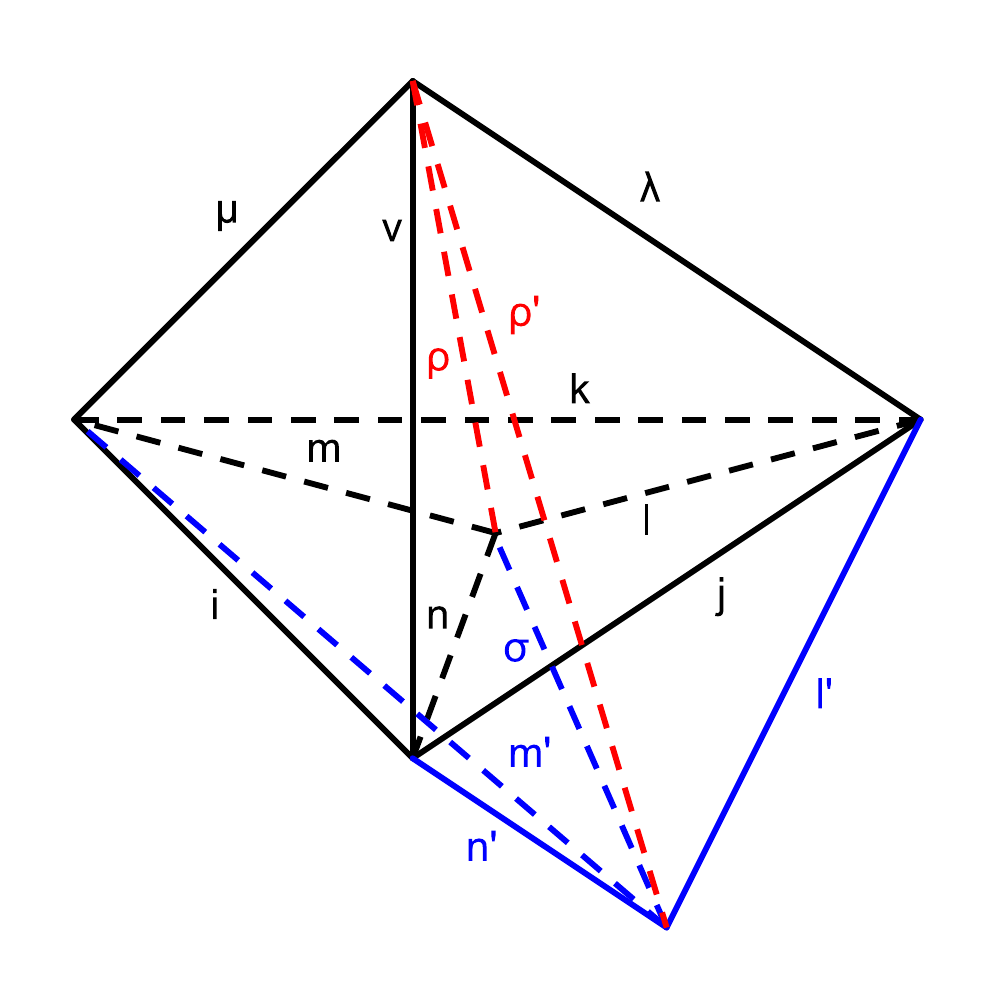}
	\caption{(Color online.) Action of $B_p^\sigma$ operator corresponds to gluing another three tetrahera.}
	\label{fig:6Tetrahedra}
\end{figure}

Gluing another three tetrahedra to the above picture, as depicted in Fig.\ref{fig:6Tetrahedra}, corresponds to fusing another loop $\sigma$ into the triangular plaquette. In the tensor network picture, $i,j,k$ remains the same, while $l,m,n,\rho$ change into $l',m',n',\rho'$. This entanglement-producing procedure of fusion can be identified as the operator $B_p^\sigma$ with matrix elements
\begin{align}
\label{eq:Bps}
&\Biggl\langle
\TriangleYpart{k}{l^{\prime}}{m^{\prime}}{n^{\prime}}{j}{i}{>}{<}{>}{<}{<}{<}
\Biggr|
B_p^\sigma
\Biggl|\TriangleYpart{k}{l}{m}{n}{j}{i}{>}{<}{>}{<}{<}{<}\Biggr\rangle\nonumber\\
=&
v_{l}v_{m}v_{n}v_{l'}v_{m'}v_{n'}
G^{jl^*n}_{\sigma n'l^{\prime*}}G^{km^*l}_{\sigma l'm^{\prime*}}G^{in^*m}_{\sigma m^{\prime}n^{\prime*}},
\end{align}
This is exactly the operator that appears in Eq.\eqref{eq:Bp}. Since it is a composition of the elementary $1-3$ and $3-1$ moves, the tensor network state \eqref{eq:TensorState} built from the UFC is an eigenstate of the $B_p$ operator with eigenvalue one:
\begin{equation}
B_p \ket{\Psi} =\ket{\Psi},~~\forall~\text{plaquette}~p.
\end{equation}
Consequently, one can act the $B_p$ operators multiple times while keeping the fixed-point tensor network states invariant. As discussed above, action of such $B_p$ operators on the tensor network states corresponds to gluing tetrahedra in the third dimension, thus the action of multiple $B_p$ operators on the same plaquette would correspond to the growth of a ``tower''.

\begin{figure}[htbp]
	\centering
	\includegraphics[scale=.32]{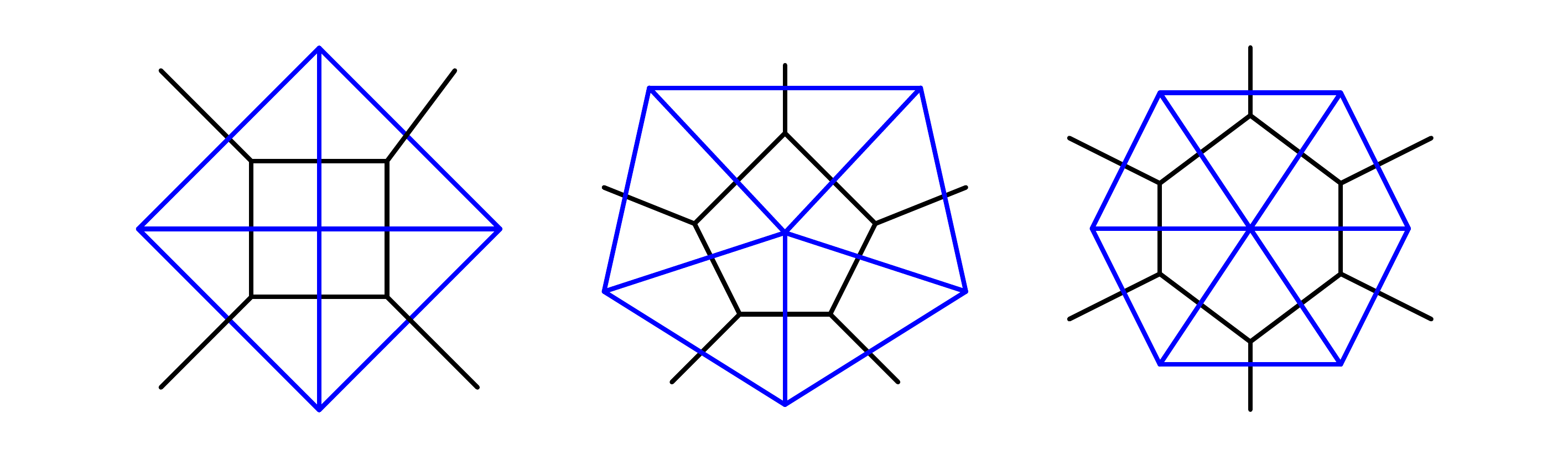}
	\caption{(color online) Plaquettes surrounded by $n=4, 5, 6$ links are drawn in black. The blue lines describe the dual triangulation picture. The auxiliary $\mu, \nu, \lambda, \cdots$ degrees of freedom has been suppressed.}
	\label{fig:lattices}
\end{figure}

Generally we can have plaquettes surrounded by more than $n\geq 3$ links, as depicted in Fig.\ref{fig:lattices}. The action of $B_p$ operators on such a plaquette would correspond to the growth of $n$ tetrahedra.

\begin{figure}[htbp]
	\centering
	\includegraphics[scale=.9]{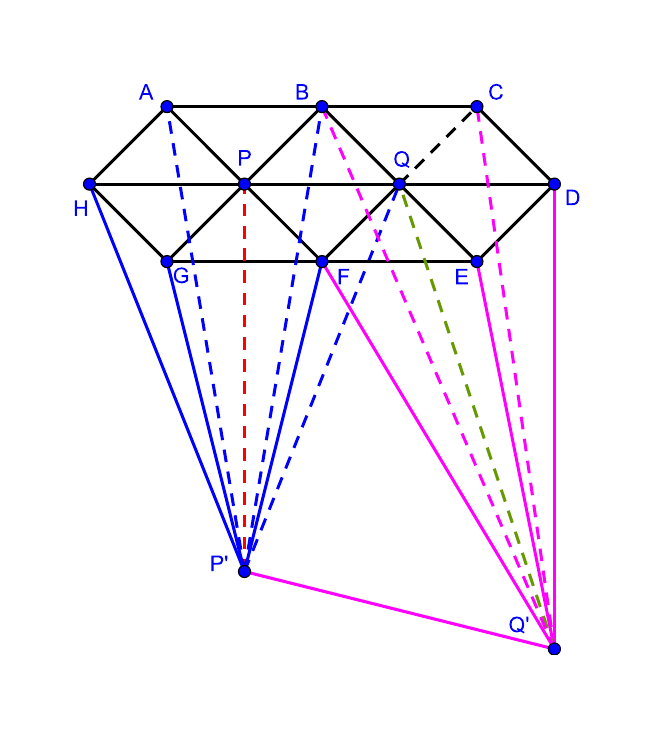}
	\caption{Action of $B_{\sigma_R}$ on the right plaquette after the action of $B_{\sigma_L}$ on the left of the plaquette.}
	\label{fig:neighbors}
\end{figure}

To illustrate the consequence of action of $B_{\sigma_L}$ and $B_{\sigma_R}$ operators on neighboring plaquettes in the tensor network picture, we consider the following honeycomb lattice as an example. In the triangulation picture (Fig.\ref{fig:neighbors}), the two neighboring plaquettes share two triangles $\triangle BPQ$ and $\triangle FPQ$. Action of $B_{\sigma_L}$ on the left plaquette corresponds to dragging $P$ to $P'$, connecting $P'$ to the six vertices of the hexagon $ABQFGH$ and thus generating six tetrahedra. Then the action of $B_{\sigma_R}$ follows, dragging $Q$ to $Q'$. One still connects $Q'$ to the five vertices $B, C, D, E, F$, but not $P$, for $P$ has already been dragged to $P'$ by the previous action of $B_{\sigma_L}$. Therefore the last line to connect would be $P'Q'$. In this way, we again generate six tetrahedra $BCQQ'$, $CDQQ'$, $DEQQ'$, $EFQQ'$, $FPP'Q'$ and $P'BQQ'$ and there is no space left unfilled, i.e., we have obtained an emergent pre-geometry in the third dimension.

It is a pre-geometry in the sense of tiling: it contains vertices that correspond to points in spacetime and oriented edges connecting them. The stronger concept of emergent bulk \textit{geometry} would further require a metric defined from measures of entanglement in the fixed-point tensor network states on the boundary.

This is exactly the implication of the \textit{Holographic Principle}[\onlinecite{Hooft,Susskind,Maldacena}], where the information in the 3D bulk is fully stored in the 2D tensor network.


We have thus shown the correspondence between fixed-point tensor network states and TQFTs in a higher dimension. If the 2D manifold where the tensor network lives in is closed, i.e., has no boundary, then degeneracy of the ground states in the topological phase described by the fixed-point tensor network state can be expressed in terms of $B_p$ operators[\onlinecite{YT1502}]. To be specific, we have 
\begin{equation}
GSD_{\mathcal{C}}(\Sigma)=tr\left(\prod_p \sum_\sigma \frac{d_\sigma}{D} B_p^\sigma\right)=\tau_{\mathcal{C}}\left(\Sigma\times S^1\right),
\end{equation}
once the action of $B_p^\sigma$ operators is identified with the gluing of three new tetrahedra that share an edge $\sigma$.

A side-remark: The universality classes of state-sum TQFTs are characterized by the Drinfeld center[\onlinecite{Etingof}] $\mathcal{Z}(\mathcal{C})$ of the UFC $\mathcal{C}$. In the context of tensor networks, the fixed-point states constructed from two different sets of data $\mathcal{C}_1$ and $\mathcal{C}_2$ which satisfy $\mathcal{Z}(\mathcal{C}_1)\sim\mathcal{Z}(\mathcal{C}_2)$ as braided tensor categories (i.e. $\mathcal{C}_1$ is Morita equivalent to $\mathcal{C}_2$[\onlinecite{Etingof}]) are ground states of the same physical phase. 

\section{Symmetry Enriched Case}\label{Sec:Symmetry}
In this section we provide an extension of the above framework when a global symmetry $\mathcal{G}$ is present. For simplicity, we take $\mathcal{G}$ to be finite, onsite, and unitary. 

To start with, we review some mathematical terminology. Given the input data of category $\mathcal{C}$, one can follow the procedure of previous sections \ref{Sec:Algebraic} and \ref{Sec:Geo} to construct a tensor network state. This state is the ground state of a topological phase described by the Drinfeld center $\mathcal{Z}(\mathcal{C})$ of $\mathcal{C}$. 

It is known[\onlinecite{Barkeshli2014}] that a large subset of $\mathcal{G}$-symmetry enriched topological phases (SETs) can be described by a braided $\mathcal{G}$-crossed extension of $\mathcal{Z}(\mathcal{C})$. To obtain such a phase, we need to use another UFC $\mathcal{D}$ as the input data of the tensor network instead of $\mathcal{C}$. This $\mathcal{D}$ is called a ``$\mathcal{G}$-extension of $\mathcal{C}$''[\onlinecite{Etingof2009}]. It is endowed with a $\mathcal{G}$-graded structure in the following way:
\begin{equation}
\label{eq:D}
\mathcal{D}=\bigoplus_{g\in \mathcal{G}} \mathcal{D}_g.
\end{equation}
Writing $e$ as the identity element of $\mathcal{G}$, we require $\mathcal{D}_e=\mathcal{C}$. In other words, if the symmetry group $\mathcal{G}$ is trivial, i.e. has only one single element $e$, $\mathcal{D}$ reduces to the original category $\mathcal{C}$. Furthermore, we require the fusion rules in $\mathcal{D}$ to be compatible with the group structure of $\mathcal{G}$. This amounts to requiring
\begin{equation}
\label{eq:fusionD}
\mathcal{D}_g\otimes\mathcal{D}_h\subset\mathcal{D}_{gh},~~~a_g\otimes b_h=\bigoplus_c N_{abc^*} c_{gh}.
\end{equation}

If we demand $\mathcal{D}$ to be the input data for the tensor network, namely, if we require the labels $i,j,k,\mu,\nu,\lambda,\cdots$ in Fig.\ref{fig:AB-Vidal} to all belong to the label set $I_\mathcal{D}$ of $\mathcal{D}$, then the tensor network state will be the ground state of a ``gauged'' model of the $\mathcal{G}$-SET in question. In such model the global symmetry $\mathcal{G}$ is promoted to a gauge symmetry and the $g\in\mathcal{G}$ fluxes become deconfined excitations of the ``gauged'' model. 

To return to the $\mathcal{G}$-SET, one has to go through an ``ungauging'' procedure[\onlinecite{Levin1606,Cheng1606}]. Since all the labels $i,j,k,\cdots,\mu,\nu,\lambda,\cdots$ belong to some $\mathcal{D}_g$, they are related by construction to a group element of $\mathcal{G}$ obtained by the map $\mathcal{D}_g \mapsto g$. For convenience, we recall the triple-line structure Fig.\ref{fig:AB-Vidal} of tensor network below. 

\begin{figure}[htbp]
\centering
\includegraphics[]{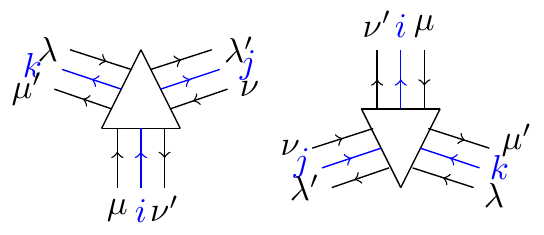}
\caption{(Color online) Triple-line structure of vertices that belong to $A$ (left) and $B$ (right) sublattices.}
\label{fig:AB-Vidal-2}
\end{figure}

We then define another group element $\tilde{g}$ for the degrees of freedom on the blue links in the left figure above:
\begin{equation}
\label{eq:tilde}
\tilde{g}_i= g_\mu^{-1}g_{\nu'},~~\tilde{g}_j= g_{\lambda'}^{-1}g_\nu,~~\tilde{g}_k=g_{\mu'}^{-1}g_\lambda.
\end{equation}
The conventions are fixed in the following way: starting from the blue link $i$ in the left figure above, rotate it 90 degrees counterclockwise. The head of the rotated $i$ link points to $\mu$ and the tail of the $i$ link joins to the head of the $\nu'$ arrow. Now invert $g_\mu$, but keep $g_{\nu'}$ unchanged. Similarly, rotate the link $j$ by 90 degrees counterclockwise. The head of the $j$ link points to $\lambda'$ and its tail to $\nu$, and so we invert $g_{\lambda'}$ but keep $g_\nu$ unchanged. One observes that there is a gauge degree of freedom in the above definition: if an arbitrary group element $g\in\mathcal{G}$ is left-multiplied to all $\mu,\mu',\nu,\nu',\lambda,\lambda',\cdots$, the definition of $\tilde{g}_{i},\tilde{g}_{j},\tilde{g}_{k}$ will remain exactly the same.

To complete the ``ungauging'' procedure, we set
\begin{equation}
\label{eq:ungauging}
g_i=\tilde{g}_i
\end{equation}
for all blue-link degrees of freedom $i,j,k$, etc.. The corresponding tensor network state is the ground state of a $\mathcal{G}$-SET on a sphere, with the tensors on each of the sublattices of the honeycomb lattice given by
\begin{widetext}
\begin{equation}
\label{eq:G-SET}
\begin{split}
& \text{A:}\ (\mathbf{T}^M)^{ijk}_{\mu\mu'\nu\nu'\lambda\lambda'}=\frac{\left(v_{\mu}v_{\nu}v_{\lambda}\right)^{1/3}}{\sqrt{D}}\sqrt{v_iv_jv_k}G^{ij^*k^*}_{\lambda\mu^*\nu}\delta_{\mu\mu'}\delta_{\nu\nu'}\delta_{\lambda\lambda'} \delta_{g_i\tilde{g}_i}\delta_{g_j\tilde{g}_j}\delta_{g_k\tilde{g}_k},\\
& \text{B:}\ (\mathbf{T}^M)^{ijk}_{\mu\mu'\nu\nu'\lambda\lambda'}=\frac{\left(v_{\mu}v_{\nu}v_{\lambda}\right)^{1/3}}{\sqrt{D}}\sqrt{v_iv_jv_k}G^{i^*jk}_{\lambda\mu^*\nu}\delta_{\mu\mu'}\delta_{\nu\nu'}\delta_{\lambda\lambda'} \delta_{g_i\tilde{g}_i}\delta_{g_j\tilde{g}_j}\delta_{g_k\tilde{g}_k},\\
\end{split}
\end{equation}
\end{widetext}
borne in mind that all the labels belong to the label set of $\mathcal{D}$, {\it not} of $\mathcal{C}$. Except for this, we notice that the form of tensors in Eq.\eqref{eq:G-SET} are the same as the previous Eq.\eqref{eq:TN-Tensors}, only with an additional flatness constraint on the $\mathcal{G}$ gauge field. The $\mathbf{T}^M$ tensors take such a simple form because $\mathcal{G}$ is onsite and unitary. A formulation for more general symmetries is possible and is related to the work in Ref.[\onlinecite{Cheng1606}]. 

The symmetry $\mathcal{G}$ manifests itself as the invariance of the tensor network state 
under a global action of $U^g$, where $U_g$ is defined as
\begin{equation}
U^g:~ g_\mu\rightarrow g_\mu g~~\forall \mu,~~g_i\rightarrow g^{-1}g_ig~~\forall i.
\end{equation}
That is, $U_g$ acts as right-multiplication by $g$ for all the group elements associated with the plaquette degrees of freedom, and acts as conjugation by $g$ for all the group elements associated with the links.

On the mathematical side, the TQFT that incorporates the $\mathcal{G}$-symmetry is known as Homotopy Quantum Field Theory (HQFT), which was proposed by Turaev[\onlinecite{Homotopy2D, Homotopy3D,TuraevBook}]. HQFT is a version of TQFT defined on some $\mathcal{G}$-manifold $\Sigma$, which is a manifold endowed with a $\mathcal{G}$ gauge field, i.e. a homotopy class of maps $\Sigma\rightarrow B\mathcal{G}$ from the manifold to the classifying space $B\mathcal{G}$. 
For connected manifolds $\Sigma$, homotopy classes of such maps correspond bijectively to the set of homomorphisms ${\rm Hom}(\pi_1(\Sigma),\mathcal{G})$, which in turn completely determine[\onlinecite{Dijkgraaf-Witten}]  principle $\mathcal{G}$-bundles over $\Sigma$. 

From an algebraic perspective, any braided $\mathcal{G}$-crossed extension of $\mathcal{Z}(\mathcal{C})$ gives rise to a HQFT with target space $B\mathcal{G}$[\onlinecite{Homotopy3D}]. Physically, every realization of symmetry $\mathcal{G}$-enriched topological phase is described by a HQFT with target space $B\mathcal{G}$.

The related symmetry-enriched TV-invariant can be constructed following Refs.[\onlinecite{Barrett1996,Homotopy3D}]. The formulation is exactly parallel to that of the tensor network above. For a triangulation of a 3D manifold with boundary, we first assign oriented labels in $I_\mathcal{D}$ to the 1-simplices of the triangulation. For a given homotopy class of maps $\Sigma\rightarrow B\mathcal{G}$, we then choose a representative map $\mathbf{g}$ that sends all the vertices of the triangulation to a base point of $B\mathcal{G}$. We then assign to each 1-simplex a group element in $\mathcal{G}$: $\mu\mapsto g_\mu$, $\mu^*\mapsto g_\mu^{-1}$.
Similar to the constraint of Eq.\eqref{eq:ungauging}, we then impose the flatness condition for all 2-simplices in the bulk of the triangulation. In other words, we require there to be no local $\mathcal{G}$-symmetry fluxes. 
Importantly, we further require the assignments $\mu\mapsto g_\mu$ to be compatible with the $\mathcal{G}$-grading stucture of $\mathcal{D}$. Namely, we require the group element $g_\mu$ assigned to edge $\mu$ to be such that $\mu\in I_{\mathcal{D}_{g_\mu}}$.

After performing this construction, one obtains the TV-invariant $\tau(\Sigma)$ of the $\mathcal{G}$-manifold $\Sigma$ in a way similar to the case without symmetry [\onlinecite{Barrett1996}]. Since Pachner moves can be extended naturally to the symmetry-enriched case, one can readily prove that $\tau(\Sigma)$ is independent of the chosen triangulation. Furthermore, $\tau(\Sigma)$ is also independent of the choice of representative $\mathbf{g}$ in the homotopy class of classifying maps [\onlinecite{Homotopy3D}].

\section{Summary and Outlook}\label{Sec:Summary}

We have identified the algebraic structure of 2D fixed-point tensor network states as unitary fusion categories, which are also known as representation theories of $C^*$-weak Hopf algebras. We illustrated how the pattern of long-range entanglement of fixed-point tensor network states arises in such a picture. 

Geometrically, we demonstrated how to construct a 2D fixed-point tensor network state from a 3D state-sum topological quantum field theory. The long-range entangled fixed-point tensor network state lives on the 2D boundary of the 3D TQFT, and encodes the same amount of information as the latter, which is a characteristic of holography. Furthermore, we showed how the emergence of bulk pre-geometry arises from the long-range entanglement of the fixed-point tensor network states on the boundary. We further extended the correspondence when a finite unitary symmetry is present.

The correspondence between the data of fixed-point tensor network states, 3D state-sum TQFT and unitary fusion category is summarized in the following table.
\begin{widetext}
	\begin{table}[htbp]
		\centering
		\begin{tabular}{|c|c|c|c|c|}
			\hline
			{\bf State Sum TQFT} &  {\bf Tensor Network} & {\bf Unitary Fusion Category} & {\bf Entanglement} \\ \hline
			Edges on the brane boundary & Interal d.o.f.s in the plaquettes & $\mu,\nu,\lambda,... \in I$. & Mutually un-entangled partons \\ 
			\hline		
			Edges on the cut boundary & Physical d.o.f.s on the links & $i,j,k,... \in I$ & Entangled physical d.o.f.s\\
			\hline
			Faces (triangles) & Triple-line strucutre & $(\mu,i,\nu),(\nu,j,\lambda),(\lambda,k,\mu),...$  & Projections $\mu\otimes \nu\rightarrow i$ etc.\\ 
			\hline
			Tetrahedra & Vertices and tensors $(T^M)^{ijk}_{\mu\nu\lambda}$ & $G^{ijk}_{\mu\nu\lambda},...$ & Carriers of entanglement\\ 
			\hline
			Invariance under Pachner move & Invariance under RG & Pentagon \& Orthogonality  & Long-range Entanglement\\
			\hline
		\end{tabular}
		\label{Tab:Sum}
	\end{table}
\end{widetext}

One future direction would be to define the fixed-point tensor network states in terms of the more familiar language of algebras, rather than categories. Namely, instead of isomorphism classes of simple objects in $\mathcal{C}$, we could use basis elements of a $C^*$-weak Hopf algebra $\mathcal{W}$ as the link labels of the trivalent graph. This requires application of the Tannakian duality $\mathcal{C}\simeq Rep(\mathcal{W})$, see for example Ref.[\onlinecite{Ostrik}]. In the simplest case of finite groups, this duality has a simple interpretation as a generalized ``Fourier transformation" [\onlinecite{Fourier}]. The more interesting quantum group cases, however, requires additional care. This idea is closely related to the work[\onlinecite{Bultinck2015}], where the authors constructed tensor network states using matrix product operators and a $C^*$ algebra. Topological phases are then described by the central idempotents of the corresponding $C^*$ algebra. 

Another extension would be to relax the Tetrahedral symmetry of the $6j$-symbols in our formulation. This could lead to interesting physics, and in the finite group case may allow us to obtain a tensor-network representation of the Dijkgraaf-Witten model.

One can go beyond the ground state subspace as well. The structure of fixed-point tensor network states that are excited states of a topological phase is expected to be characterized by a TQFT with marked surfaces.


\begin{acknowledgments}
We thank Meng Cheng, Ling-Yan Hung, Alex Turzillo and Zhao Yang for very helpful suggestions on the manuscript. We are grateful to Dominic J. Williamson for informing us of his previous work. Zhuxi appreciates the elucidating conversations with Brendan Pankovich and Fei Teng.
\end{acknowledgments}

\appendix
\section{3D Pachner Moves and Consistency Conditions for the $G$-tensors}
\label{App:Pachner}
In this appendix, we sketch how the Pentagon equation \eqref{eq:UFC-Pentagon} and Orthogonality condition \eqref{eq:UFC-Ortho} are related to the three dimensional Pachner moves. 

\begin{figure}[htbp]
\centering
\includegraphics[]{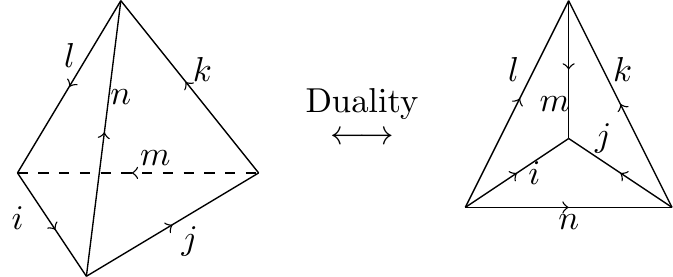}
\caption{Faces of the tetrahedron are mapped to the vertices of the dual diagram, edges are mapped to links, and vertices are mapped to triangles.}
\label{fig:3DPoincare}
\end{figure}

Every 3-simplex in the 3D triangulation can be mapped[\onlinecite{Muger}] to a two dimensional categorical diagram. Here faces of the tetrahedron are mapped to the vertices of the dual diagram, edges are mapped to links, and vertices are mapped to triangles. When two tetrahedra in the triangulation share a face, the dual diagram possesses two triangles which share a vertex. In these cases we draw the diagrams separately and connect the common vertex with a dashed line. 


These 2D diagrams naturally inherit an action of the Pachner moves from the 3D triangulation. Algebraically, since every 3-simplex is directly related to a $6j$-symbol (as illustrated in Sec.\ref{SubSec:GeoTQFT}), so is every 2D categorical diagram. Below we show that algebraic expressions for the invariance under Pachner moves in the categorical diagrams are exactly the Pentagon and Orthogonality constraints.

\begin{widetext}
\begin{figure}[h]
\centering
\includegraphics[]{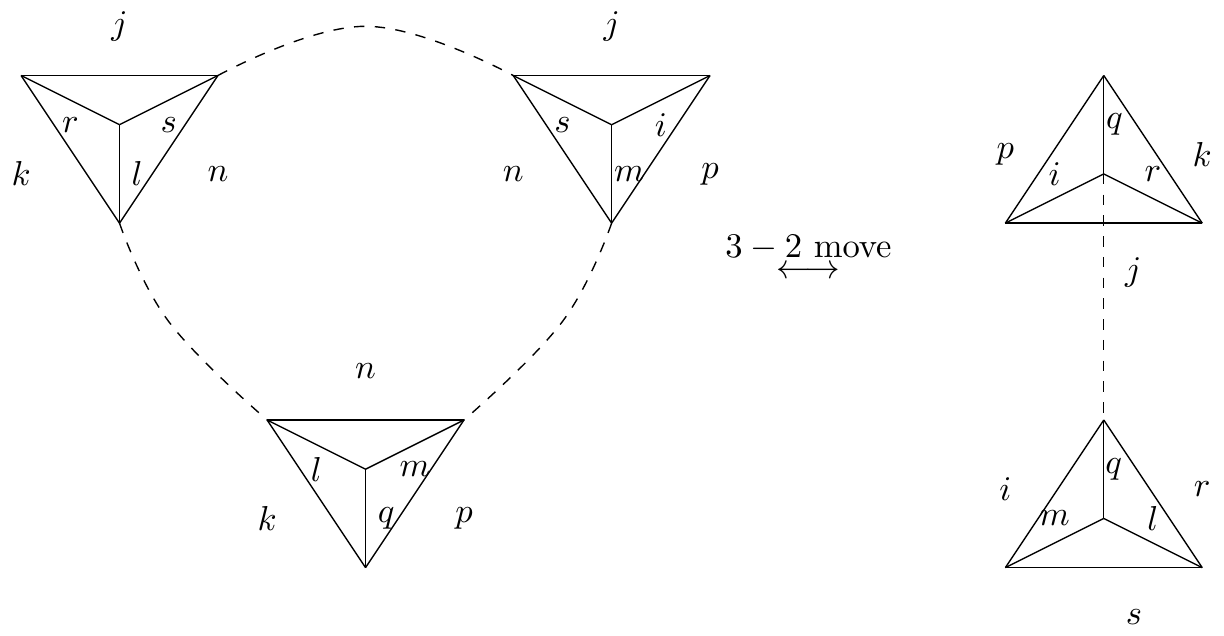}
\caption{3D Pachner 3-2 move.}
\label{fig:3DPachner3-2}
\end{figure}
\end{widetext}

\begin{figure}[h]
\centering
\includegraphics[]{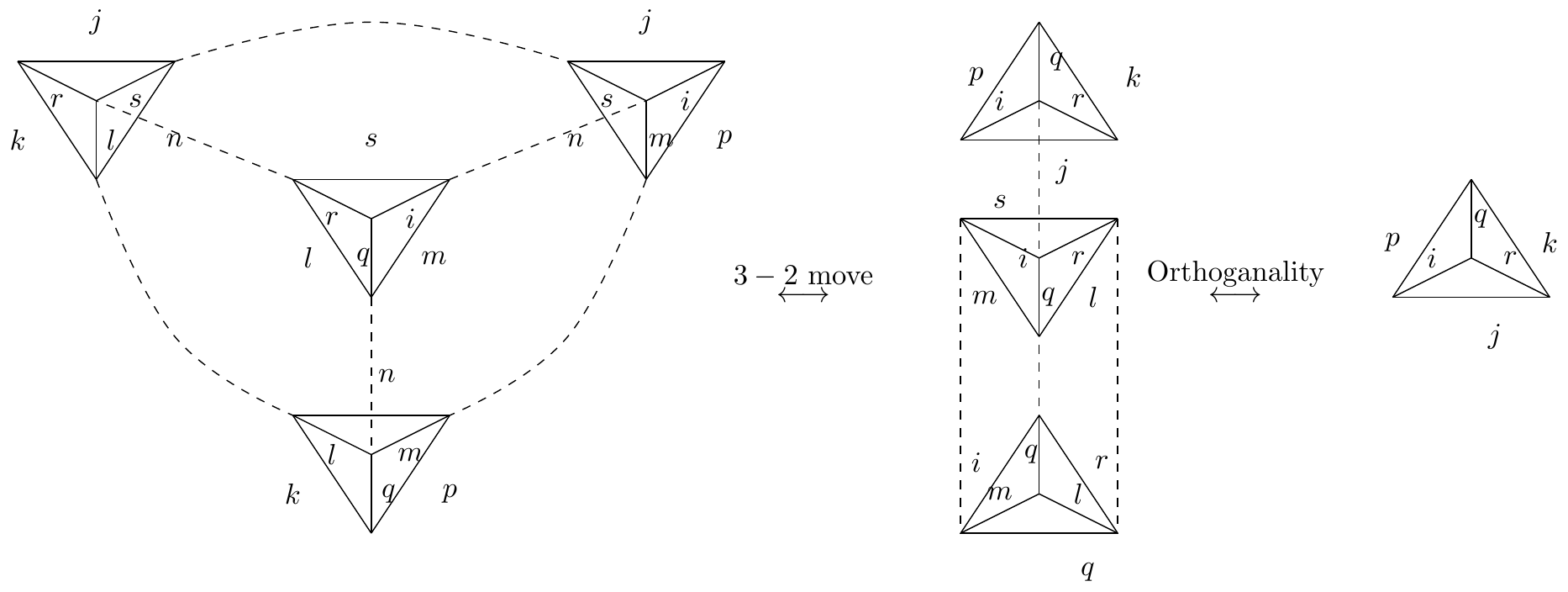}
\caption{3D Pachner 4-1 move.}
\label{fig:3DPachner4-1}
\end{figure}


\clearpage

\begin{thebibliography}{50}

\bibitem{Chen2010}
X.~Chen, Z.-C.~Gu, and X.-G.~Wen, {\it Local unitary transformation, long-range quantum entanglement, wave function renormalization, and topological order}, Phys. Rev. B 82, 155138 (2010).

\bibitem{OrdusReview}
R.~Or{\'u}s, {\it A practical introduction to tensor networks: Matrix product states and projected entangled pair states}, Ann. Phys. 349, 117–158 (2014).
		
\bibitem{Meets}
B.~Zeng, X.~Chen, D.-L.~Zhou and X.-G. Wen, {\it Quantum information meets quantum matter--from quantum entanglement to topological phase in many-body systems}, {ArXiv preprint:1508.02595}.
		
\bibitem{MPS1991}
A.~Klumper, A.~Schadschneider and J.~Zittartz, {\it Equivalence and solution of anisotropic spin-1 models and generalized tJ fermion models in one dimension}, J. Phys. A 24, 16 (1991).

\bibitem{MPS1992}
M.~Fannes, B.~Nachtergaele and R.~F. Werner, {\it Finitely correlated states on quantum spin chains}, Commun. Math. Phys. 144, 443 (1992).

\bibitem{MPS1993}
A.~Kl{\"u}mper , A.~Schadschneider and J.~Zittartz, {\it Matrix product ground states for one-dimensional spin-1 quantum antiferromagnets},  Europhys. Lett. 24, 293 (1993).

\bibitem{PEPS}
F.~Verstraete and J.~I. Cirac, {\it Renormalization algorithms for quantum-many body systems in two and higher dimensions}, {ArXiv preprint: cond-mat/0407066}.

\bibitem{Swingle}
B.~Swingle, {\it Entanglement renormalization and holography}, {Phys. Rev. D}~Phys. Rev. D 86, 6 (2012).

\bibitem{Evenbly2011}
G.~Evenbly and G.~Vidal, {\it Tensor network states and geometry},~ J. Stat. Phys. 145, 4 (2011).

\bibitem{QiEHM}
X.-L.~Qi, {\it Exact holographic mapping and emergent space-time geometry}, {ArXiv preprint: 1309.6282}.

\bibitem{HaPPY}
F.~Pastawski, B.~Yoshida, D.~Harlow and J.~Preskill, {\it Holographic quantum error-correcting codes: Toy models for the bulk/boundary correspondence}, {ArXiv preprint: 1503.06237}.

\bibitem{YZ2015}
Z.~Yang, P.~Hayden and X.-L. Qi, {\it Bidirectional holographic codes and sub-AdS locality}, {ArXiv preprint: 1510.03784}.

\bibitem{YZ2016}
P.~Hayden, S.~Nezami, X.-L. Qi, N. Thomas, M.~Walter and Z.~Yang, {\it Holographic duality from random tensor networks}, {ArXiv preprint: 1601.01694}.

\bibitem{Hooft}
G.~t'Hooft, {\it Dimensional reduction in quantum gravity}, {ArXiv preprint: gr-qc/9310026}.

\bibitem{Susskind}
L.~Susskind, {\it The world as a hologram}, {J. Math. Phys.}~36, 11 (1995).

\bibitem{Maldacena}
J.~Maldacena, {\it The large-N limit of superconformal field theories and supergravity}, Adv. Theor. Math. Phys. 2, 231 (1998).

\bibitem{Rovelli1995}
C.~Rovelli and L.~Smolin, {\it Spin networks and quantum gravity}, {Phys. Rev. D}~52, 10 (1995).

\bibitem{Rovelli1998}
C.~Rovelli, {\it Loop quantum gravity}, {Living Rev. Rel}~1, 1 (1998).

\bibitem{Barrett1998}
J.~W. Barrett and L.~Crane, {\it Relativistic spin networks and quantum gravity}, {J. Math. Phys.}~39, 6 (1998).

\bibitem{Han2016}
M.-X. Han and L.-Y. Hung, {\it Loop quantum gravity, exact holographic mapping, and holographic entanglement entropy}, {ArXiv preprint: 1610.02134}.
	
\bibitem{AKLT}
I.~Affleck, T.~Kennedy, E.~H.~Lieb and H.~Tasaki, {\it Rigorous results on valence-bond ground states in antiferromagnets}, {Phys. Rev. Lett.}~ 59, 7 (1987).

\bibitem{Atiyah}
M.~F. Atiyah, {\it Topological quantum field theory}, Publ. Math. IHES 68, 175–186 (1989).

\bibitem{TV}
V.~G. Turaev and O.~Y. Viro, {\it State sum invariants of 3-manifolds and quantum 6j-symbols}, {Topology}~ 31, 4 (1992).

\bibitem{Barrett1996}
J.~Barrett and B.~Westbury, {\it Invariants of piecewise-linear 3-manifolds}, {Trans. Amer. Math. Soc.}~ 348, 10 (1996).

\bibitem{Kapustin1607}
A.~Kapustin, A.~Turzillo and M.-Y. You, {\it Topological field theory and matrix product states}, {ArXiv preprint: 1607.06766}.

\bibitem{Ryu1607}
K.~Shiozaki and S.~Ryu, {\it Matrix product states and equivariant topological field theories for bosonic symmetry-protected topological phases in (1+ 1) dimensions}, {ArXiv preprint:1607.06504}.

\bibitem{Singh2010}
S.~Singh, R.~N. C.~Pfeifer and G.~Vidal, {\it Tensor network decompositions in the presence of a global symmetry}, {Phys. Rev. A}~ 82, 5 (2010).

\bibitem{Ran2015}
S.-H. Jiang and Y.~Ran, {\it Symmetric tensor networks and practical simulation algorithms to sharply identify classes of quantum phases distinguishable by short-range physics}, {Phys. Rev. B}~92, 10 (2015).

\bibitem{Homotopy2D}
V.~G. Turaev, {\it Homotopy field theory in dimension 2 and group-algebras}, {ArXiv preprint: math/9910010}.

\bibitem{Homotopy3D}
V.~G. Turaev, {\it Homotopy field theory in dimension 3 and crossed group-categories}, {ArXiv preprint: math/0005291}.

\bibitem{TuraevBook}
V.~G. Turaev, {\it Homotopy quantum field theory}, {Euro. Math. Soc.}~ 10, (2010).

\bibitem{Levin1606}
C.~Heinrich, F.~Burnell, L.~Fidkowski and M. A. Levin, {\it Symmetry enriched string-nets: Exactly solvable models for SET phases}, {Phys. Rev. B} 94, 23 (2016).

\bibitem{Cheng1606}
M.~Cheng, Z.-C.~Gu, S.-H.~Jiang and Y.~Qi, {\it Exactly solvable models for symmetry-enriched topological phases}, {ArXiv preprint:1606.08482}.

\bibitem{Dijkgraaf-Witten}
R.~Dijkgraaf and E.~Witten, {\it Topological gauge theories and group cohomology}, {Commun. Math. Phys.}~129, 2 (1990).

\bibitem{TQD}
Y.-T.~Hu, Y.-D.~Wan and Y.-S.~Wu, {\it Twisted quantum double model of topological phases in two dimensions}, {Phys. Rev. B}~87, 12 (2013).

\bibitem{Gu2009}
Z.-C.~Gu, M.~A.~Levin, B.~Swingle and X.-G.~Wen, {\it Tensor-product representations for string-net condensed states}, {Phys. Rev. B}~79, 8 (2009).

\bibitem{Buerschaper2009}
O.~Buerschaper, M.~Aguado and G.~Vidal, {\it Explicit tensor network representation for the ground states of string-net models}, {Phys. Rev. B}~79, 8 (2009).

\bibitem{Lan2014}
T.~Lan and X.-G. Wen, {\it Topological quasiparticles and the holographic bulk-edge relation in (2+ 1)-dimensional string-net models}, {Phys. Rev. B}~90, 11 (2014).

\bibitem{TERG}
Z.-C.~Gu, M.~A.~Levin and X.-G.~Wen, {\it Tensor-entanglement renormalization group approach as a unified method for symmetry breaking and topological phase transitions}, {Phys. Rev. B}~78, 20 (2008).

\bibitem{Evenbly2015}
G.~Evenbly and G.~Vidal, {\it Tensor network renormalization}, {Phys. Rev. Lett.}~115, 18 (2015).

\bibitem{Evenbly2016}
G.~Evenbly and G.~Vidal, {\it Local scale transformations on the lattice with tensor network renormalization}, {Phys. Rev. Lett.}~116, 4 (2016).

\bibitem{MacLane}
S.~MacLane, {\it Categories for the working mathematician}, {Springer Science \& Business Media}~5, (2013).

\bibitem{LW}
M.~A.~Levin and X.-G.~Wen, {\it String-net condensation: A physical mechanism for topological phases}, {Phys. Rev. B}~71, 4 (2005).

\bibitem{Pachner}
U.~Pachner, {\it PL homeomorphic manifolds are equivalent by elementary shellings}, {Euro. J. Combin.}~12, 2 (1991).

\bibitem{Etingof}
P.~Etingof, S.~Gelaki, D.~Nikshych and V.~Ostrik, {\it Tensor categories}, {Amer. Math. Soc.}~205, (2015).

\bibitem{Karowski1992}
M.~Karowski, W.~Muller and R.~Scharder, {\it State sum invariants of compact 3-manifolds with boundary and 6j-symbols}, {J. Phys. A}~25, 18 (1992).

\bibitem{YT1502}
Y.-T.~Hu, N.~Geer and Y.-S. Wu, {\it Full dyon excitation spectrum in generalized levin-wen models}, {ArXiv preprint: 1502.03433}.


\bibitem{Barkeshli2014}
M.~Barkeshli, P.~Bonderson, M.~Cheng and Z.-H. Wang, {\it Symmetry, defects, and gauging of topological phases}, {ArXiv preprint: 1410.4540}.

\bibitem{Etingof2009}
P.~Etingof, D.~Nikshych and V.~Ostrik, {\it Fusion categories and homotopy theory}, {ArXiv preprint: 0909.3140}.

\bibitem{Ostrik}
V.~Ostrik, {\it Module categories, weak hopf algebras and modular invariants}, {Trans. Grps.}~8, 2 (2003).

\bibitem{Fourier}
O.~Buerschaper and M.~Aguado, {\it Mapping kitaev’s quantum double lattice models to Levin and Wen’s string-net models}, {Phys. Rev. B}~ 80, 15 (2009).

\bibitem{Bultinck2015}
N.~Bultinck, M.~Mari{\"e}n, D.~J. Williamson, M. B. {\c{S}}ahino{\u{g}}lu, J.~Haegeman and Frank F.~Verstraete, {\it Anyons and matrix product operator algebras}, {ArXiv preprint: 1511.08090}.

\bibitem{Muger}
M.~M{\"u}ger, {\it On TQFTs}, (2007).

\bibitem{Kirillov}
A.~Kirillov Jr., {\it String-net model of Turaev-Viro invariants}, {ArXiv preprint: 1106.6033}.

\bibitem{Koenig}
R.~Koenig, G.~Kuperberg and B.~W. Reichardt, {\it Quantum computation with Turaev-Viro codes}, {Ann. Phys.}~325, 12 (2010).

\bibitem{Wang1606}
D.~J. Williamson and Z-H. Wang, {\it Hamiltonian realizations of (3+1)-TQFTs}, {ArXiv preprint:1606.07144}.

\end{thebibliography}
\end{document}